\documentclass[sigplan,nonacm]{acmart}

\AtBeginDocument{%
  \providecommand\BibTeX{{%
    Bib\TeX}}}

\setcopyright{none} 
\copyrightyear{2025}
\acmYear{2025}
\acmDOI{XXXXXXX.XXXXXXX}

\usepackage{quantikz}
\usepackage{algorithmic}
\usepackage{amsmath}
\usepackage{caption}
\usepackage{subcaption}
\usepackage{soul}
\usepackage{xspace}
\usepackage{float}
\usepackage{graphicx}
\usepackage{algorithm,multirow}
\usepackage{grffile}
\usepackage{makecell}
\usepackage{enumitem}
\usepackage{url}
\usepackage{hyperref}

\newcommand{\sol}{InterPlace}

\def\BibTeX{{\rm B\kern-.05em{\sc i\kern-.025em b}\kern-.08em
    T\kern-.1667em\lower.7ex\hbox{E}\kern-.125emX}}

\begin{document}

\title{Optimizing Inter-chip Coupler Link Placement for Modular and Chiplet Quantum Systems}

\author{
Zefan Du$^{1}$,
Pedro Chumpitaz Flores$^{2}$,
Wenqi Wei$^{1}$,
Juntao Chen$^{1}$,
Kaixun Hua$^{2}$,
Ying Mao$^{1}$\\
$^{1}$Fordham University, New York, NY, USA\\
$^{2}$University of South Florida, Tampa, FL, USA\\
\{zdu19, wwei23, jchen504, ymao41\}@fordham.edu, \{pedrochumpitazflores, khua\}@usf.edu
}

\begin{abstract}

Quantum computing offers unparalleled computational capabilities but faces significant challenges, including limited qubit counts, diverse hardware topologies, and dynamic noise/error rates, which hinder scalability and reliability. Distributed quantum computing, particularly chip-to-chip connections, has emerged as a solution by interconnecting multiple processors to collaboratively execute large circuits. While hardware advancements, such as IBM's Quantum Flamingo~\cite{ibmFlamingo2024}, focus on improving inter-chip fidelity, limited research addresses efficient circuit cutting and qubit mapping in distributed systems. This project introduces \sol, a self-adaptive, hardware-aware framework for chip-to-chip distributed quantum systems. \sol~analyzes qubit noise and error rates to construct a virtual system topology, guiding circuit partitioning and distributed qubit mapping to minimize SWAP overhead and enhance fidelity. Implemented with IBM Qiskit and compared with the state-of-the-arts, \sol~achieves up to a 53.0\% improvement in fidelity and reduces the combination of on-chip SWAPs and inter-chip operations by as much as 33.3\%, demonstrating scalability and effectiveness in extensive evaluations on real quantum hardware topologies.

\end{abstract}

\maketitle

\pagestyle{plain}

\section{Introduction}

Quantum computing is approaching a turning point. Declared by the United Nations as the International Year of Quantum Science and Technology in 2025~\cite{iyq}, the field is entering a phase where theoretical breakthroughs must translate into practical, scalable systems. Among the most promising use cases is quantum chemistry for drug discovery, where simulating molecular interactions at quantum accuracy offers transformative potential beyond classical limits~\cite{cao2019quantum}, early applications have emerged from different fields, such as machine learning~\cite{stein2022quclassi,ullah2024quantum,stein2021hybrid,liu2024training,stein2021qugan,verma2025quantum, dai2025quantum,l2024quantum}, quantum-classical high performance computing~\cite{kan2024scalable, elsharkawy2025integration, mu2022iterative, du2024efficient, d2023distributed}, and privacy/security~\cite{li2025quantum,abdikhakimov2024preparing, namakshenas2024federated, qu2024quantum, baseri2024navigating}.

Superconducting quantum hardware, currently a leading platform, has achieved remarkable progress in gate fidelity, coherence time, and device scale. Yet, as monolithic chips expand, they confront critical scaling bottlenecks: limited fabrication yields, dense routing constraints from control wiring, thermal inefficiencies, and manufacturing complexity. These constraints have led to a growing consensus: the future of scalable quantum systems lies in modular architectures, where smaller quantum chips are interconnected to act as a unified processor~\cite{monroe2014large, nickerson2014freely, du2025hardware, du2024hardware}.

Modularity offers multiple advantages. It enhances fabrication feasibility, simplifies thermal and wiring management, and allows for localized error correction. This trend is gaining momentum—Google’s Willow system showcased high-fidelity inter-chip gates via tunable couplers \cite{blogMeetWillow}, and IBM’s Flamingo architecture outlines a scalable cryogenic layout for chip-to-chip coupling \cite{ibmFlamingo2024}. Other quantum platforms, such as trapped ions and neutral atoms, are similarly pursuing modularity via photonic or teleportation-based links. However, this introduces system-level challenges, particularly in deciding which qubits should host inter-chip couplers, as different placements lead to varying error rates and potential congestion on critical links. While prior efforts have focused on hardware-level coupler designs or software-level compiler optimizations, these layers often operate in isolation, leaving few frameworks that jointly optimize physical connectivity and system-level performance. 


\begin{figure}[]
    \centering
    \includegraphics[width=0.9\linewidth]{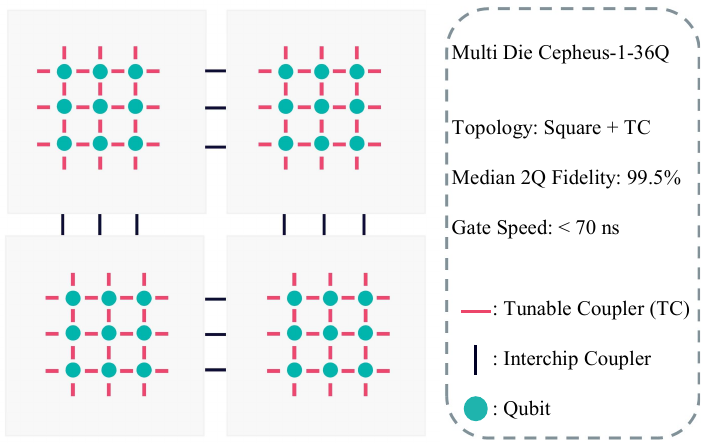}
    \caption{Rigetti's multi-chip Architecture. The Cepheus-1-36Q multi-chip system connects four 9-qubit chip with inter-chip couplers~\cite{ankaa}}.
    \label{fig:Rigetti}
    \vspace{-0.2in}
\end{figure}

This paper proposes \sol, a Inter-chip coupler link placement framework. It addresses the gap through a hardware–software co-design framework that tightly integrates physical layout constraints with system-level performance objectives. Central to our approach is an integrated cost model for selecting $k$ inter-chip coupler links between two or more quantum chips, balancing trade-offs among latency, congestion, and hardware feasibility. Importantly, our model is applicable to both modular and chiplet-based systems, since its parameters can be tuned to reflect different link characteristics: modular couplers typically span longer distances, while chiplet couplers are much shorter, such as modular quantum system from IBM-Q~\cite{ibmFlamingo2024} and chiplet quantum systems from Rigetti as shown in Figure~\ref{fig:Rigetti}. 

\sol~ formulate a cost model that balances routing efficiency, realistic communication quality, system-level reliability, congestion, and manufacturability. The formulation penalizes overloaded, high-degree qubits while favoring low-error, low-latency, spatially balanced connections, yielding configurations resilient to routing inefficiencies and hardware limits. Optimizing \sol's cost at manufacturing phase produces inter-chip connections that minimize inter-chip communication delay subject to constraints such as maximum qubit degree and coupler density. This enables a feedback loop between architectural planning and hardware layout, guiding early-stage decisions that affect long-term performance and error resilience.
To make the model practical, we introduce a greedy selection workflow that efficiently evaluates link candidates across real and synthetic topologies. 
Our contributions are summarized below:

\begin{itemize}
    \item We introduce \sol, a hardware-aware framework that jointly optimizes inter-chip coupler placement and system-level communication. This co-design approach bridges physical feasibility with logical efficiency, addressing a gap left by hardware-only or compiler-only methods.
    \item \sol~ employs a multi-objective cost model that integrates fidelity-aware latency, routing efficiency, congestion, qubit degree limits, and spatial distribution. The cost model provides a metric for evaluating inter-chip connections before fabrication.
    \item We design an efficient optimization procedure that scales across large chip topologies and outputs inter-chip coupler placements fully compatible with existing compilers.
    \item We implement \sol~ in Python/Qiskit and evaluate it with 5 state-of-the-art compilers, including Qiskit, Cirq, Pytket, MQT, and UCC. Across diverse workloads and system configurations, InterPlace consistently reduces SWAP overhead and inter-chip operations, yielding up to 53.0\% fidelity improvement and 33.3\% fewer On-chip SWAPs and Inter-chip operations compared to baseline link placements. 
    
\end{itemize}

\section{Related Work}

The pursuit of scalable quantum computing has increasingly turned toward modular architectures to overcome the limitations of monolithic processors. Designs such as an all-to-all reconfigurable router~\cite{WuModular} and MIT’s large-scale integration platform~\cite{MIT2024} highlight this potential but largely overlook systematic handling of inter-chip placement. Industry roadmaps now emphasize modularity: IBM’s processors, such as Crossbill with M-coupler and Flamingo with L-coupler, integrate short- and long-range couplers to scale qubit communication~\cite{IBMResearch2025AnnualLetter, ibmFlamingo2024}. These developments foreground the need to plan link endpoints and densities alongside package and layout choices, so that connectivity is provisioned where traffic and calibration constraints are expected to concentrate. In this context, inter-chip placement becomes a first-order architectural concern rather than a post hoc packaging decision.

Effective inter-chip links are central. Cavity-mediated interconnects~\cite{BenRached2024} and photonic shuttling devices~\cite{MIT2025} show promise but face fidelity, bandwidth, and calibration limits that interact with placement and spacing rules. Photonic platforms also demonstrate scale—Xanadu’s 35-chip cluster-state system with 86.4 billion modes illustrates the feasibility of large-scale modular photonics~\cite{Xanadu2025}. At the device level, performance characterization of static inter-chip couplers informs operating windows and coupling targets relevant to multi-chip layouts~\cite{Norris2025}. Together, these results frame the trade-offs among latency, fidelity-aware delay, spatial clustering, and manufacturability that arise when selecting and positioning inter-chip couplers.

On the software side, qubit mapping and routing are critical for minimizing inter-module overhead under fixed device graphs. QUBO-based graph partitioning reduces costly inter-core operations~\cite{Bandic2023}, while toolchains like SEQC~\cite{Jeng2025} improve placement and routing but often treat hardware limits as post-processing. Our approach instead embeds these constraints directly into the link-selection process so that routing pressure, degree limits, and spacing considerations are reflected in the chosen inter-chip links. Other works explore error mitigation and dynamic circuits; real-time classical links enable processors to operate as a unified unit~\cite{Vazquez2024}. In aggregate, these studies underscore the importance of modular architectures, inter-chip coupling mechanisms, and optimization frameworks in advancing scalable quantum computing systems.

\color{blue}

\color{black}


\section{Background and Motivation}


\subsection{Essential Physical Qubit Information}

A physical qubit is defined not only by its logical behavior but also by physical characteristics that govern reliability and efficiency in quantum computation. Understanding these parameters is essential for hardware-aware compilation, architecture co-design, and system-level optimization.
In superconducting devices, key metrics capture coherence, control performance, and physical attributes. The most fundamental coherence properties are the relaxation time $T_1$, which measures how long an excited qubit remains before decaying to $\ket{0}$, and the dephasing time $T_2$, which quantifies how long a superposition such as $\ket{+}$ maintains phase coherence~\cite{kjaergaard2020superconducting}. These parameters determine the effective lifetime of stored quantum information.

At the gate level, two categories of metrics are distinguished. For single-qubit 
operations such as $X$, $H$, or $R_z$, performance is captured by the error rate 
$\epsilon_{\text{1Q}}$ and the gate time $t_{\text{1Q}}$. For two-qubit operations, 
which are typically the most error-prone, the relevant measures are the error rate 
$\epsilon_{\text{2Q}}$ and execution time $t_{\text{2Q}}$~\cite{arute2019quantum}. Since 
algorithms often rely on repeated entangling gates, these two-qubit metrics usually 
dominate overall circuit fidelity. Qubit measurement (or readout) introduces additional 
constraints. The accuracy is limited by a readout error probability $\epsilon_{\text{readout}}$, 
while the speed is determined by the readout time $t_{\text{readout}}$~\cite{barends2014superconducting}. 
Reliability is further affected by cross-talk, denoted $\epsilon_{\text{xtalk}}$, 
which represents the chance of an error occurring on a qubit due to simultaneous activity 
on neighboring qubits. Even when idle, qubits accumulate idle errors, primarily 
dictated by their $T_1$ and $T_2$ values~\cite{gambetta2017building}.


Beyond error and timing, devices include topological and physical metadata: a unique qubit identifier; physical coordinates $(x,y,z)$ on the chip; and the set of directly connected neighbors supporting native two-qubit gates. Additional properties influencing compilation and calibration include operating frequency (with spectral crowding constraints), anharmonicity (for selective control of computational states), thermal population (probability of being thermally excited into $\ket{1}$), and tunability (available frequency adjustment in flux-controlled qubits)~\cite{krantz2019quantum}. Together, these properties define the operational envelope of a quantum device and inform mapping and optimization decisions.




\subsection{Chip Topology and Qubit Connectivity}
\begin{figure}[H]
    \centering
    \begin{subfigure}[t]{0.22\linewidth}
        \centering
        \includegraphics[width=\linewidth]{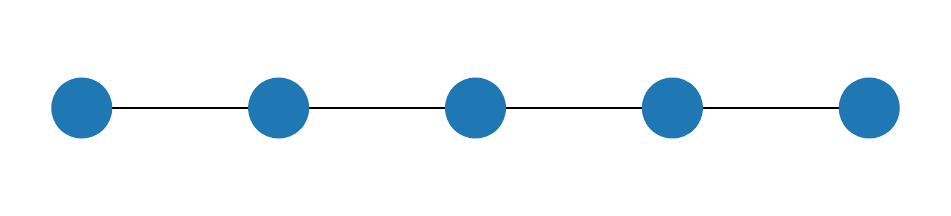}
        \caption{Line}
    \end{subfigure}\hfill
    \begin{subfigure}[t]{0.22\linewidth}
        \centering
        \includegraphics[width=\linewidth]{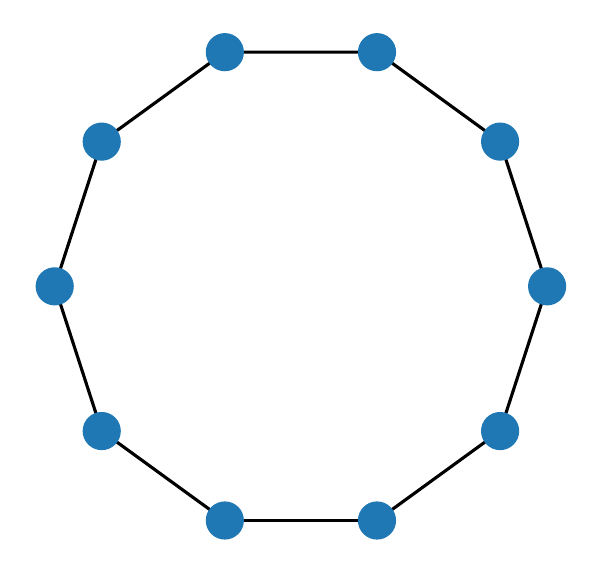}
        \caption{Ring}
    \end{subfigure}\hfill
    \begin{subfigure}[t]{0.22\linewidth}
        \centering
        \includegraphics[width=\linewidth]{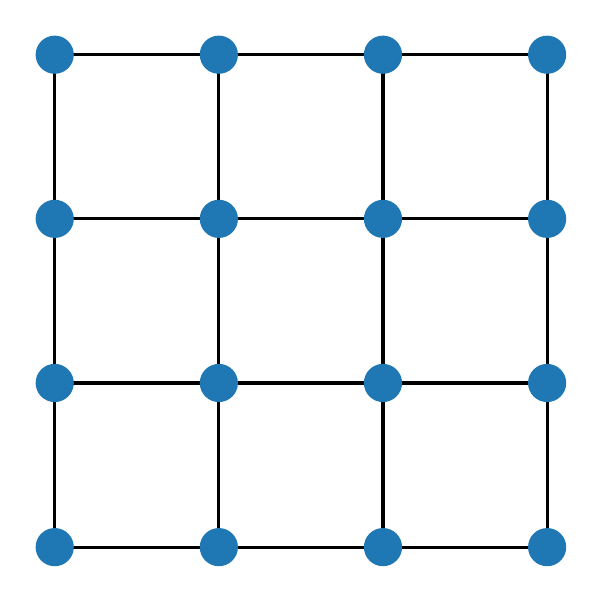}
        \caption{Grid}
    \end{subfigure}\hfill


    \begin{subfigure}[t]{0.22\linewidth}
        \centering
        \includegraphics[width=\linewidth]{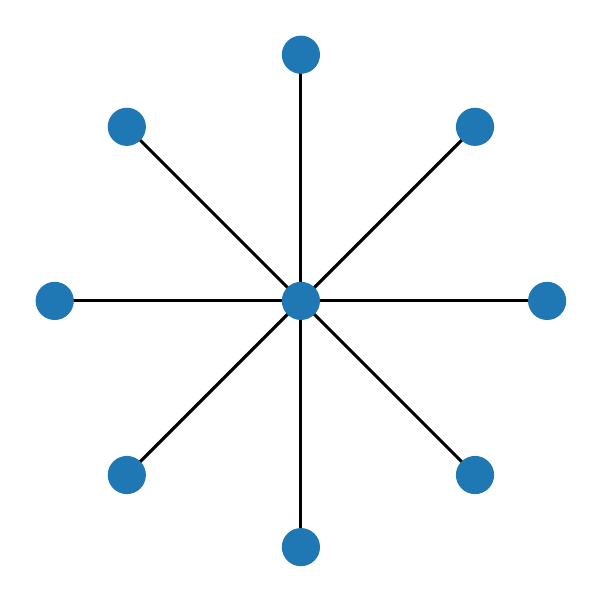}
        \caption{Star}
    \end{subfigure}\hfill
    \begin{subfigure}[t]{0.22\linewidth}
        \centering
        \includegraphics[width=\linewidth]{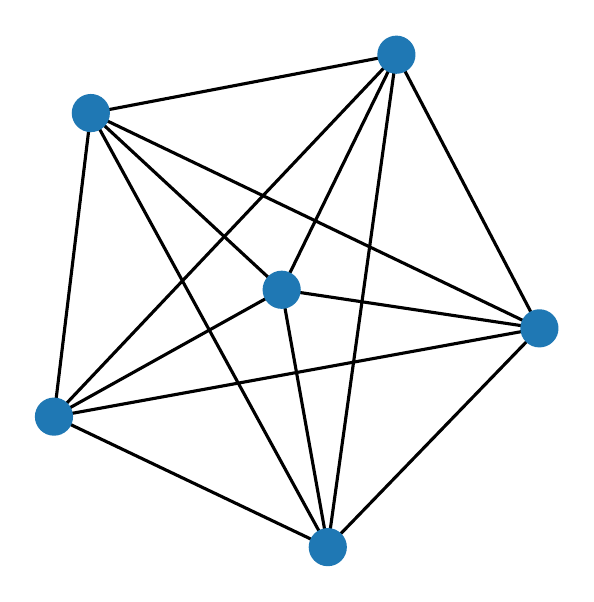}
        \caption{Complete}
    \end{subfigure}\hfill
    \begin{subfigure}[t]{0.22\linewidth}
        \centering
        \includegraphics[width=\linewidth]{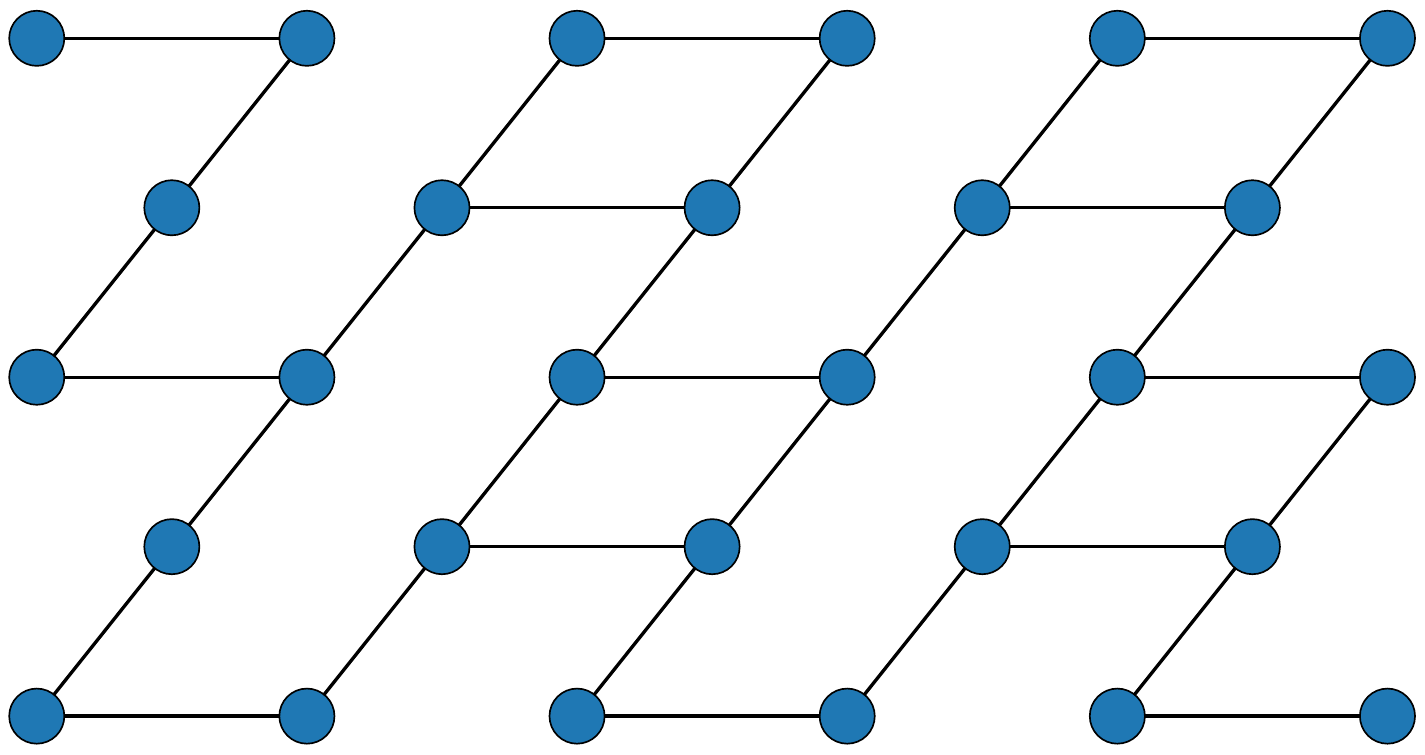}
        \caption{Heavy-Hex}
    \end{subfigure}

    \caption{Representative qubit connectivity graphs.}
    \vspace{-0.1in}
    \label{fig:topologies}
\end{figure}

Qubit mapping and routing are governed by the processor’s connectivity graph. 
Figure~\ref{fig:topologies} shows six canonical patterns—\emph{Line}, \emph{Ring}, \emph{Grid}, \emph{Star}, \emph{Complete (all-to-all)}, and \emph{Heavy-Hex} layout representative of IBM devices. 
Degree distribution and graph diameter shape shortest-path distances and congestion; combined with calibrated edge error rates, they determine the effective time-to-fidelity of multi-qubit interactions. 
For instance, a Star minimizes path length but concentrates traffic at a hub; Rings equalize load but increase average distance; Grids and Ladders trade diameter for layout regularity; Heavy-Hex bounds degree ($\le 3$) to mitigate crosstalk while preserving routability. 
These structural properties motivate the cost terms and coupling strategies used in this work.

\subsection{Multi-Chip Architectures \& Inter-Chip Coupling}

In multi-chip quantum systems, inter-chip qubit communication is enabled by physical couplers that connect qubits from different modules. These couplers exhibit variability in fidelity, latency, and physical footprint. For instance, IBM's m-couplers support short-range high-fidelity links, while l-couplers enable long-distance communication at the expense of performance~\cite{ibmFlamingo2024}. Physically, couplers are limited by layout constraints and fabrication tolerances. They occupy chip area, add thermal load, and require precise calibration. Logically, poorly chosen coupler links can increase circuit depth and require additional SWAP operations, compounding decoherence and gate errors. Additionally, overuse of individual couplers can create performance bottlenecks. Hence, effective inter-chip coupling must minimize communication latency while ensuring balanced load distribution, physical feasibility, and logical efficiency.

Multi-chip quantum architectures aim to overcome the limitations of monolithic scaling by interconnecting smaller chips via physical couplers~\cite{monroe2014large, nickerson2014freely}. These modules can be fabricated and tested independently, facilitating parallel development and higher yield. Recent systems like Google’s Willow~\cite{Google2024Willow} and IBM’s Crossbill and Flamingo~\cite{IBMResearch2025AnnualLetter, ibmFlamingo2024} exemplify this approach. They use dedicated couplers, including medium- and long-range types, embedded within multi-layer chip stacks. While modular designs simplify fabrication and cooling, they introduce new optimization challenges related to communication overhead, spatial congestion, and interconnect reliability. The fundamental trade-off lies in balancing connectivity and fidelity — more couplers improve flexibility but increase resource contention and complexity.

\subsection{Motivation for Our Work}


Current methods for optimizing circuit layout and compilation often focus on logical-level placement and routing without fully incorporating physical constraints. For example, QUBO, a model of qubit placement, minimizes logical communication but assume a fixed hardware topology~\cite{Bandic2023}. More advanced tools like SEQC~\cite{Jeng2025} enable scalable backend compilation for modular devices, but still rely on fixed coupler maps determined post-fabrication. These approaches are reactive—adapting software to hardware after design, rather than proactive, where hardware is co-designed with logical requirements in mind. At the hardware level, design heuristics based on empirical layouts are used, but lack formal performance models to justify coupler placement. In practice, most optimization occurs at the compiler and transpilation phases, while at the hardware layer there is still no effective cost model to rigorously quantify the quality of specific coupler link choices. Consequently, there remains a disconnect between physical constraints and logical optimization.

Designing modular and chiplet quantum hardware requires aligning physical layout with logical communication requirements. Treating these aspects independently can produce long routing paths, concentrated traffic, and increased error rates. We formulate inter-chip coupler selection as a co-design problem that integrates hardware constraints with communication objectives. The evaluation of candidate couplers is based on five criteria: (i) routing distance, (ii) latency–fidelity trade-off, (iii) local link congestion, (iv) qubit degree limits, and (v) spatial distribution. These are formally defined in Section~\ref{sec:problem_formulation}. This framework supports the analysis of inter-chip connectivity on both real and synthetic device models before fabrication, enabling systematic exploration of modular and chiplet quantum architectures.

\section{\sol~ Cost Modeling} \label{sec:problem_formulation}

\sol~ models the design of inter-chip coupler placement in multi-chip quantum systems as a constrained network design problem. 
Given two processing modules, each with an internal network of interconnected nodes, we aim to select a set of cross-module links that optimize end-to-end communication efficiency while respecting physical and operational constraints. 

In the context of \emph{modular quantum architectures}, these modules correspond to quantum chips $A$ and $B$, each containing a set of physical qubits and native on-chip couplings. 
We represent the internal connectivity of chip $A$ as $G_A = (V_A, E_A)$ and chip $B$ as $G_B = (V_B, E_B)$, where:
\begin{itemize}
    \item $V_i$: set of qubits in chip $i$, each with hardware properties $\{t_{\mathrm{coh}}, \epsilon_{\mathrm{readout}}\}$.
    \item $E_i$: set of available on-chip quantum gates, each edge weighted by $(t_e, \epsilon_e)$, representing gate duration and error probability.
\end{itemize}

\noindent Our decision variables correspond to selecting $n$ \emph{inter-chip coupler links} between candidate pairs $(u, v)$, with $u \in V_A$ and $v \in V_B$. 
The design goal is to choose these links such that the resulting modular system achieves high communication performance, robustness to errors, and physical feasibility.

\subsection{Optimization Objective}

We define a \emph{multi-objective global cost function} that evaluates the quality of the entire selected set of links. 
This function integrates both logical (graph-theoretic) and physical (hardware-level) performance measures. 
The objective is:

\begin{equation}
\begin{aligned}
\text{Total Cost} =\; & \alpha \cdot \text{Average Path Length} \\
& + \beta \cdot \text{Effective Path Cost} \\
& + \gamma \cdot \text{Congestion Penalty} \\
& + \delta \cdot \text{Qubit Overload Penalty} \\
& + \epsilon \cdot \text{Sparsity Penalty},
\end{aligned}
\label{eq:cost_fn}
\end{equation}

\noindent where $\alpha, \beta, \gamma, \delta, \epsilon \geq 0$ are tunable weights reflecting the relative importance of each criterion. 
Each term in \eqref{eq:cost_fn} is defined as: 
 \textbf{Average Path Length} — promotes central placement of coupler endpoints to minimize average on-chip routing distance, reducing SWAP overhead and logical circuit depth.
\textbf{Effective Path Cost} — incorporates both communication latency and error rates through a \emph{Time-to-Fidelity} (TTF) model, producing a fidelity-aware delay metric.
\textbf{Congestion Penalty} — discourages spatial clustering of inter-chip couplers to reduce cross-talk, thermal load, and control-line contention.
 \textbf{Qubit Overload Penalty} — limits the number of inter-chip couplers incident to any single qubit, preventing resource saturation and preserving parallel operation capacity.
 \textbf{Sparsity Penalty} — encourages spatial distribution of coupler endpoints to facilitate manufacturability and reduce localized interference.


\noindent We use \emph{congestion} to capture local capacity/load effects and \emph{sparsity} to encourage broader spatial coverage and fabrication ease. The optimization task is to select the set of $n$ inter-chip coupler links that minimizes Equation~\eqref{eq:cost_fn}, subject to hardware and layout constraints.

\subsubsection{Average Path Length}
\label{sec:path}

This metric quantifies the topological centrality of a potential coupler endpoint in its respective module. 
From a general network design perspective, it is the mean shortest-path distance from each endpoint to all other nodes in the same module. 
Low values indicate that a node can be reached in fewer hops, reducing routing overhead.
In multi-chip quantum systems, each additional hop corresponds to a SWAP operation, which adds execution time and increases the probability of decoherence and gate errors. 
For a candidate pair $(u, v)$, we define:$\text{Path}(u, v) = \frac{1}{|V_A|} \sum_{x \in V_A} d_{G_A}(u, x) + \frac{1}{|V_B|} \sum_{y \in V_B} d_{G_B}(v, y) + 1,$
where $d_{G_A}$ and $d_{G_B}$ denote on-chip shortest-path distances in chips $A$ and $B$, respectively. 
The $+1$ accounts for the inter-chip hop introduced by the coupler itself. For fixed $n$, this term adds a constant offset $n$ to the objective, so it does not affect the optimal solution, only the absolute value of the cost.
Equivalently, one may denote these distances as $d_G$ and $d_H$; here we keep $d_{G_A}$ and $d_{G_B}$ to make the chip association explicit.

\noindent The global contribution used in \eqref{eq:cost_fn} is:
\begin{equation}
\text{Average Path Length} =
\sum_{u \in V_A} \sum_{v \in V_B} x_{uv} \cdot \text{Path}(u,v),
\end{equation}
with $x_{uv} \in \{0,1\}$ indicating selection of $(u,v)$.

\subsubsection{Effective Path Cost}

While the average shortest path measures hop count, it does not capture the fact that in quantum systems, each hop has an associated fidelity cost. 
We therefore use a fidelity-aware delay model called \emph{Time-to-Fidelity} (TTF), which unifies latency and error into a single effective measure.
TTF calculation is performed as a pre-processing step before the main optimization. This ensures that all candidate link costs already reflect realistic communication delays that include the impact of gate errors, so the optimizer operates directly on physically meaningful metrics.

\noindent Formally, the total contribution of this term to the optimization objective is:
\begin{equation}
\text{Effective Path Cost} = \sum_{u \in V_A} \sum_{v \in V_B} x_{uv} \cdot \text{AvgTTF}_{\text{pair}}(u, v),
\label{eq:effective_path_cost}
\end{equation}
where $x_{uv} \in \{0,1\}$ indicates whether coupler $(u,v)$ is selected.

\noindent\textbf{On-Chip TTF Edge Weights:}  
For each chip ($A$ and $B$), we compute the shortest TTF paths between all pairs of qubits using Dijkstra’s algorithm on a weighted graph.  
The weight assigned to an edge $e$ representing a physical gate is:


\begin{equation}
\text{TTF}_{\text{edge}}(e) = t_{\text{gate}}(e) + \lambda \cdot \ln\!\left(\tfrac{1}{1 - \epsilon_{\text{gate}}(e)}\right),
\label{eq:ttf_edge}
\end{equation}
where $t_{\text{gate}}(e)$ is the gate execution time, 
$\epsilon_{\text{gate}}(e)$ its error rate, and 
$\lambda > 0$ a scaling factor weighting fidelity in the delay metric.

The logarithmic penalty models the exponential decay of fidelity with repeated application of imperfect gates, inflating the cost of unreliable operations.
\emph{Note that $\epsilon_{\text{gate}}(e)$ corresponds to the notation $\epsilon_G(e)$ used for gate error rates.}

\noindent\textbf{Average on-Chip Access/Egress Costs:}  
Let $\text{TTF}_A(a \to u)$ denote the shortest TTF from qubit $a$ to $u$ on chip $A$ (and similarly for $\text{TTF}_B$ on chip $B$). We define the following averages:
\begin{equation}
\text{TTF}_A(u) = \frac{1}{|V_A|} \sum_{a \in V_A} \text{TTF}_A(a \to u),
\end{equation}
\begin{equation}
\text{TTF}_B(v) = \frac{1}{|V_B|} \sum_{b \in V_B} \text{TTF}_B(v \to b),
\end{equation}
where $\text{TTF}_A(u)$ and $\text{TTF}_B(v)$ denote averages on all sources / sinks within each chip.

\noindent\textbf{Inter-Chip Coupler TTF:}  
The physical inter-chip coupler between $u \in V_A$ and $v \in V_B$ has its own latency and error:
\begin{equation}
\text{TTF}_{\text{coupler}}(u \to v) = t_{\text{coupler}}(u, v) + \lambda \cdot \ln\!\left(\frac{1}{1 - \epsilon_{\text{coupler}}(u, v)}\right).
\end{equation}

\noindent\textbf{Per-Pair Average TTF (precomputed):}  
Combining the above, the fidelity-aware delay for routing through a selected coupler $(u,v)$ is:
\begin{equation}
\text{AvgTTF}_{\text{pair}}(u, v) = \text{TTF}_A(u) + \text{TTF}_{\text{coupler}}(u \to v) + \text{TTF}_B(v).
\end{equation}
This quantity is precomputed for all candidate pairs and plugged directly into the \emph{global} objective via \eqref{eq:effective_path_cost}.

\subsubsection{Congestion Penalty}

When multiple inter-chip couplers are placed in close physical proximity, they may compete for the same control and readout lines, increase local wiring density, and introduce cross-talk between adjacent couplers. Such effects can degrade operational fidelity and throughput, and in extreme cases may even cause local thermal hotspots or signal interference. To discourage these configurations, we assign a penalty that grows when too many couplers are concentrated in the same region. Let $\mathcal{R}(u)$ denote the physical coordinate of qubit $u$. The congestion measure is defined as
\begin{align}
\text{Cong}(u,v) &= 
\underbrace{\text{load}(u) + \text{load}(v)}_{\text{degree penalty}} \nonumber \\
&\quad + \eta \sum_{\substack{(u',v') \in \mathcal{L} \\ (u',v') \neq (u,v)}}
\frac{1}{\text{dist}(\mathcal{R}(u),\mathcal{R}(u'))^2}.
\end{align}

where $\text{load}(q)$ is the number of couplers incident to qubit $q$, dist $(\mathcal{R}(u),\mathcal{R}(u'))$ is the physical distance between qubits on the same chip, and $\eta$ is a tunable parameter controlling the strength of the cross-talk term. The first contribution penalizes excessive reuse of individual qubits, while the second discourages spatial clustering of endpoints by assigning larger costs to nearby pairs.  
For computational efficiency in large-scale optimization, we approximate this measure by
\begin{equation}
\text{Cong}_{\text{approx}}(u, v) = \max(\text{load}(u), \text{load}(v)).
\end{equation}
The global contribution to \eqref{eq:cost_fn} is then
\begin{equation}
\text{Congestion Penalty} =
\sum_{u \in V_A} \sum_{v \in V_B} x_{uv} \cdot \text{Cong}_{\text{approx}}(u, v).
\end{equation}

\begin{figure}[H]
  \centering
  \includegraphics[width=\linewidth]{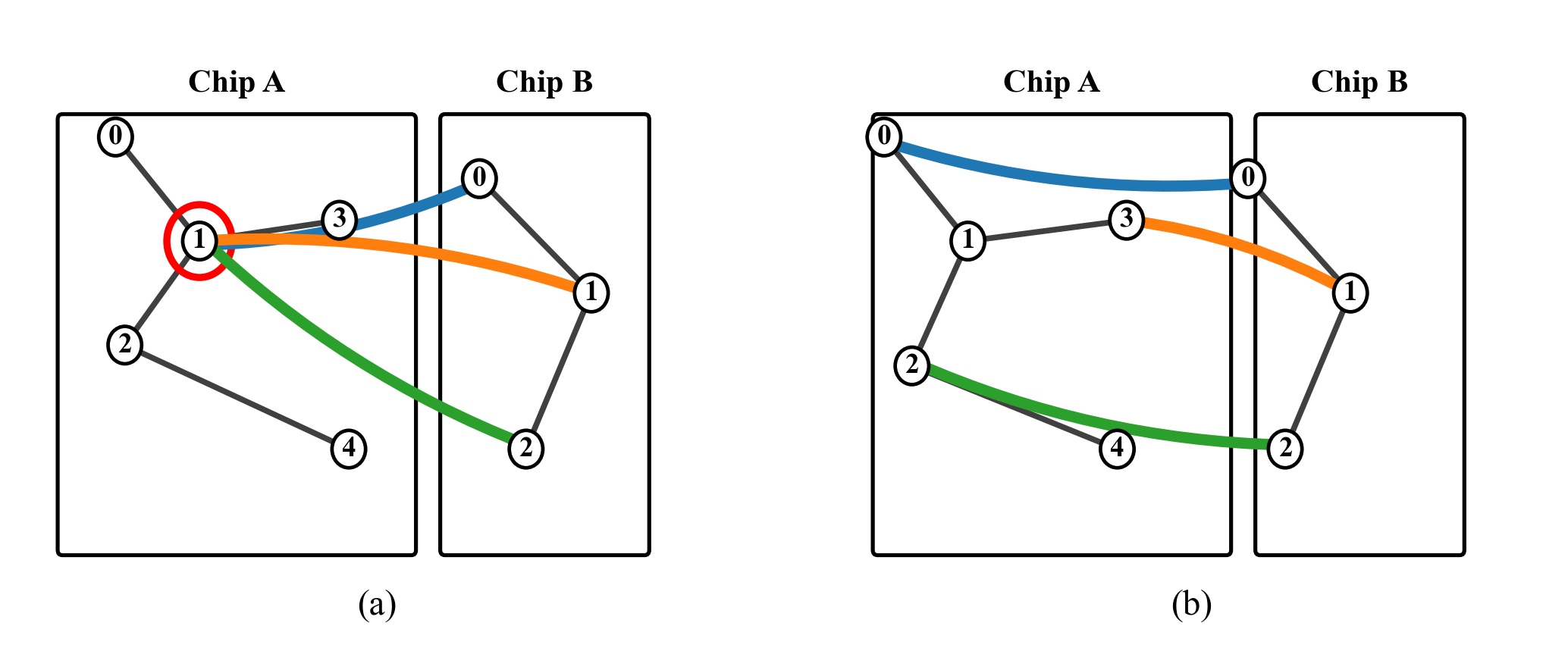}
  \caption{Coupler congestion. The left panel (a) illustrates high congestion, where multiple links share the same endpoint, leading to a larger penalty $\sum_{(u,v)\in\mathcal L}\max(\mathrm{load}(u),\mathrm{load}(v))$. The right panel (b) shows reduced congestion, where links are spread across different endpoints.}
  \label{fig:congestion}
\end{figure}

\noindent\textbf{Two-case comparison.}
Using the approximation above, consider the two configurations in Fig.~\ref{fig:congestion}.
In the left panel, the selected links are 
$\mathcal{L}_{\text{L}}=\{(1_A,0_B),(1_A,1_B),(1_A,2_B)\}$. 
The endpoint loads are $\mathrm{load}(A_1)=3$ and $\mathrm{load}(B_0)=\mathrm{load}(B_1)=\mathrm{load}(B_2)=1$. 
Hence
\begin{equation*}
\begin{aligned}
\text{penalty} 
&= \max(3,1)+\max(3,1)+\max(3,1) \\
&= 3+3+3 = 9 .
\end{aligned}
\end{equation*}
In the right panel, the links are 
$\mathcal{L}_{\text{R}}=\{(0_A,0_B),(3_A,1_B),(2_A,2_B)\}$. 
All used endpoints have $\mathrm{load}=1$, so
\begin{equation*}
\begin{aligned}
\text{penalty} 
&= \max(1,1)+\max(1,1)+\max(1,1) \\
&= 1+1+1 = 3 .
\end{aligned}
\end{equation*}
Thus, the left configuration yields a larger congestion penalty than the right configuration.



\subsubsection{Qubit Overload Penalty}

Each qubit supports a limited number of physical coupler connections, denoted by $D_{\max}$, as determined by hardware design. Exceeding this limit can saturate control and readout channels and reduce parallel operation capacity. We model this by tracking the degree $\deg(q)$ of each qubit $q$ after adding a potential link $(u,v)$, i.e., $\deg(q)+x_{uv}$ with $x_{uv}\in\{0,1\}$. If selecting $(u,v)$ causes any endpoint to exceed $D_{\max}$, a penalty is applied:
\begin{equation}
\begin{aligned}
\text{Overload Penalty} &=
\sum_{u \in V_A} \sum_{v \in V_B} x_{uv} \cdot 
\mathbb{I}\Big[ \deg(u) + x_{uv} > D_{\max} \\
&\quad \lor\ \deg(v) + x_{uv} > D_{\max} \Big].
\end{aligned}
\end{equation}

\begin{figure}[H]
  \centering
  \includegraphics[width=\linewidth]{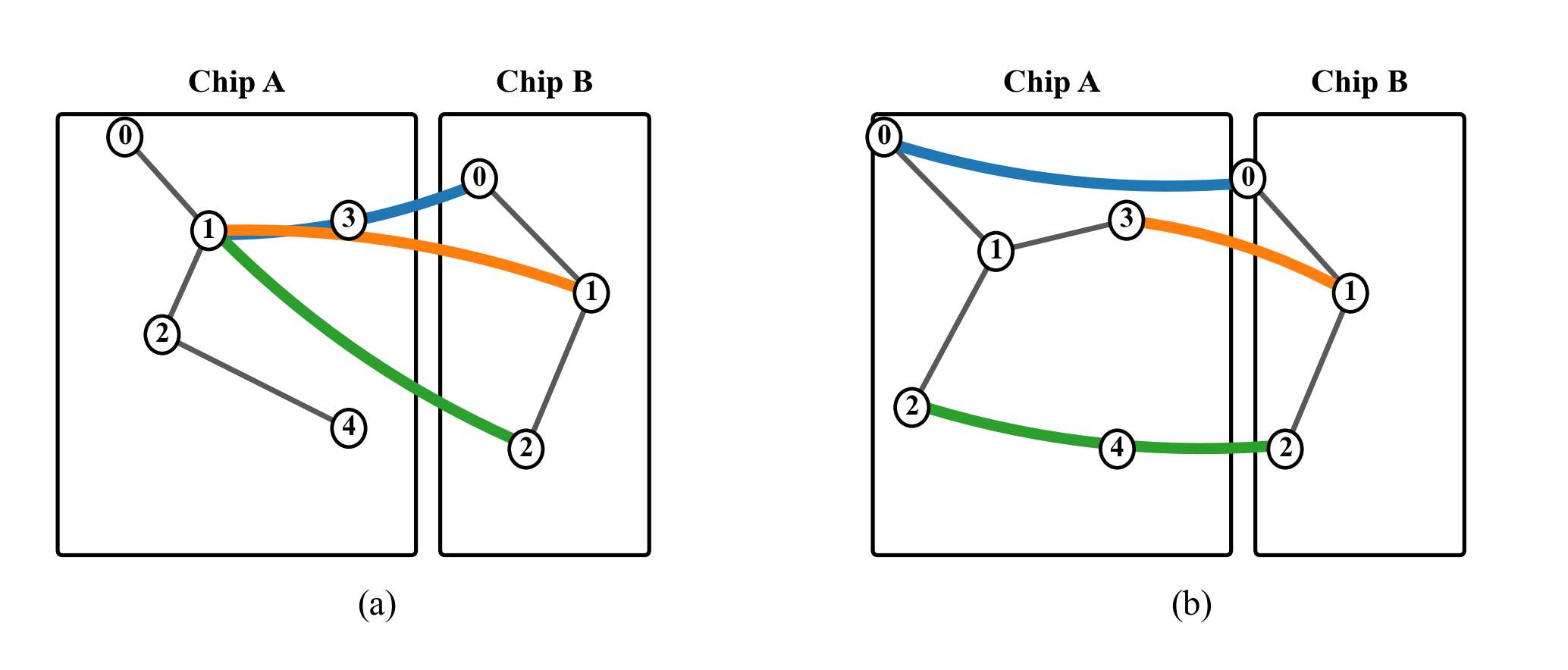}
  \caption{Qubit overload. The left panel (a) shows overload at qubit A1, where the per-qubit degree limit $D_{\max}=2$ is exceeded. The right panel (b) distributes the inter-chip couplers to respect the limit. Overload is enforced as a constraint and may also be penalized in the objective.}
  \label{fig:overload}
\end{figure}

\noindent\textbf{Two-case comparison.}
Consider the two configurations in Fig.~\ref{fig:overload} with $D_{\max}=2$. In the left panel, the selected links are $\{(1_A,0_B),$ $(1_A,1_B),(1_A,2_B)\}$, so $\deg(A_1)=3$ and $\deg(B_0)=\deg(B_1)=\deg(B_2)=1$. For each link incident to $A_1$, the quantity $\deg(A_1)$ excluding that link is $2$; hence $\deg(A_1)+x_{uv}=3>2$ and the indicator equals $1$ for all three links, so the total penalty is $1+1+1=3$. In the right panel, the links are $\{(0_A,0_B),(3_A,1_B),(2_A,2_B)\}$ and all used endpoints satisfy $\deg\le 2$, so no indicator is triggered and the penalty is $0$.

\subsubsection{Sparsity Penalty}

A high density of couplers in a small region can lead to routing congestion and fabrication challenges. To avoid this, we introduce a sparsity penalty that discourages the placement of many links in close proximity. Let $\mathcal{L}$ denote the set of selected links.  
The sparsity term in \eqref{eq:cost_fn} aggregates over unordered pairs of distinct links:
\begin{equation}
\text{Sparsity Penalty} =
\sum_{\substack{(u,v),(u',v') \in \mathcal{L} \\ (u,v) < (u',v')}} 
\frac{1}{1 + \text{dist}((u, v), (u', v'))},
\end{equation}
where 
$\text{dist}((u,v),(u',v')) = d_{G_A}(u,u') + d_{G_B}(v,v')$,  
with $d_{G_A}$ and $d_{G_B}$ the on-chip shortest path metrics defined in Section~\ref{sec:path}.  
Shorter distances lead to larger penalties, while longer distances reduce the contribution.

\begin{figure}[H]
  \centering
  \includegraphics[width=\linewidth]{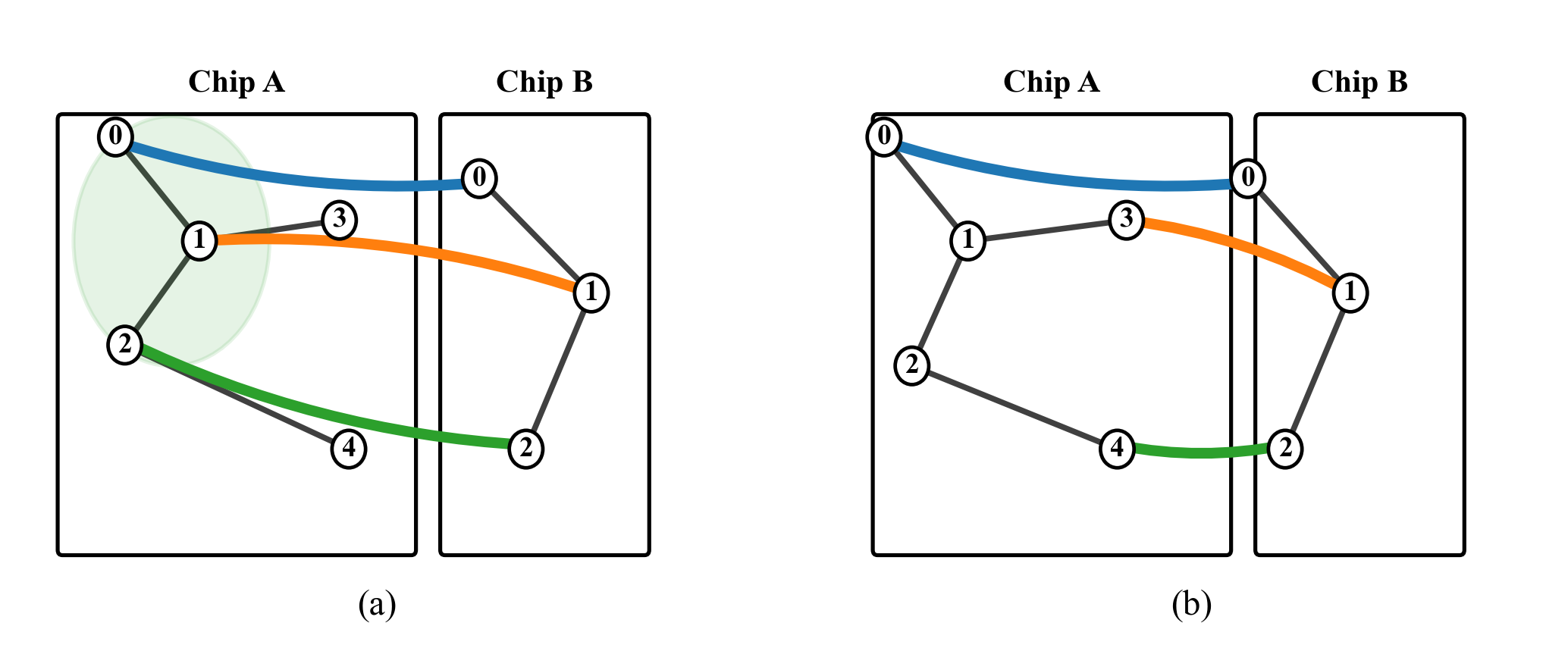}
  \caption{Spatial sparsity. The left panel (a) depicts clustered endpoints in Module~A, resulting in a high sparsity cost. The right panel (b) shows distributed endpoints across Module~A, which lowers the cost. The penalty aggregates over pairs of links using $\mathrm{dist}((u,v),(u',v')) = d_{G_A}(u,u') + d_{G_B}(v,v')$.}
  \label{fig:sparsity}
\end{figure}

\noindent\textbf{Two-case comparison.}  
To illustrate the sparsity penalty, consider the two configurations shown in Fig.~\ref{fig:sparsity}. In the first case (left panel), the links are  
$\mathcal{L}_{\text{L}}=\{(0_A,0_B), (1_A,1_B), (2_A,2_B)\}$.  
On Module~$A$, $d_{G_A}(0,1)=1$, $d_{G_A}(1,2)=1$, $d_{G_A}(0,2)=2$,  
and on Module~$B$ we have $d_{G_B}(0,1)=1$, $d_{G_B}(1,2)=1$, $d_{G_B}(0,2)=2$.  
Hence, the pairwise distances are dist((0,0),(1,1))=2, dist((0,0),(2,2))=4, dist((1,1),(2,2))=2, giving:

\begin{equation*}
\begin{aligned}
\text{penalty} &= \tfrac{1}{1+2}+\tfrac{1}{1+4}+\tfrac{1}{1+2} 
               = \tfrac{13}{15}\approx 0.867 .
\end{aligned}
\end{equation*}
In the second case (right panel), the links are  
$
\mathcal{L}_{\text{R}} =
\{(0_A,0_B),\\
  (3_A,1_B),
  (4_A,2_B)\}
$
Here $d_{G_A}(0,3)=2$, $d_{G_A}(0,4)=3$, $d_{G_A}(3,4)=3$,  
and on Module~$B$ we have $d_{G_B}(0,1)=1$, $d_{G_B}(1,2)=1$, $d_{G_B}(0,2)=2$.  
Thus, dist((0,0),(3,1)) = 3, dist((0,0),(4,2)) = 5, dist((3,1),(4,2)) = 4, yielding: 
\begin{equation*}
\begin{aligned}
\text{penalty} &= \tfrac{1}{1+3}+\tfrac{1}{1+5}+\tfrac{1}{1+4}
               = \tfrac{37}{60}\approx 0.617 .
\end{aligned}
\end{equation*}
Thus, the clustered case (left) has a higher penalty ($0.867$) than the more spread-out case (right, $0.617$).

\begin{figure*}[t]
    \centering
    \includegraphics[width=0.9\linewidth]{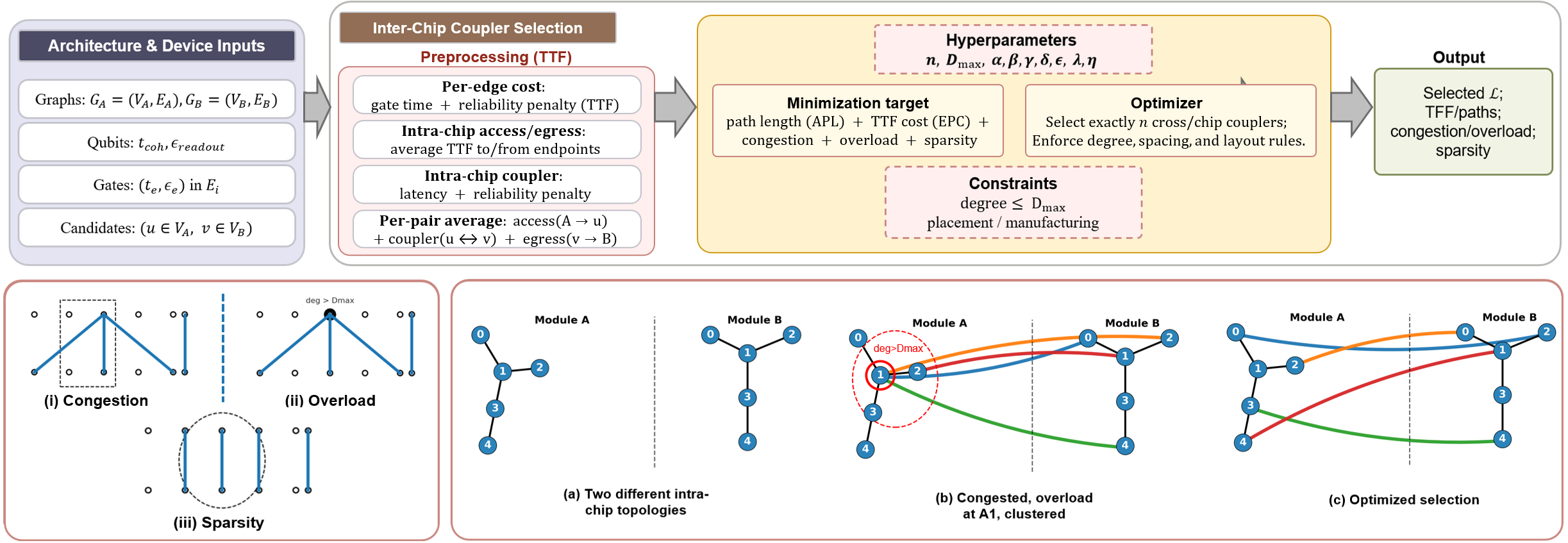}
    \caption{\sol~ Framework Overview. Top: workflow from device inputs through TTF-based preprocessing to a constrained multi-objective optimizer; output is the selected link set and metrics.Bottom: (a) two on-chip topologies; (b) random couplers induce congestion, overload (deg>$D_{\max}$), and clustering; (c) optimized couplers distribute endpoints and minimize total cost.}
    \label{fig:system_overview}
\end{figure*}

\section{\sol~ System Design}
The \sol~ cost model addresses the inter-chip coupler placement problem from multiple perspectives, such as path length, gate errors, latency, and congestion. This section introduces the \sol~ framework, which aims to minimize the cost in a multi-chip system.

\subsection{Framework Overview}
\label{sec:system_overview}

We cast the modular quantum processor as a multi-layer network in which on-chip couplings and inter-chip couplers coexist under strict hardware constraints. The central challenge is to select a subset of couplers that balances latency, fidelity, and manufacturability while avoiding local congestion or overload.  Our approach integrates the following components: (1) \textbf{Preprocessing:} latency and error rates are unified into a time-to-fidelity (TTF) cost, which allows us to precompute effective pairwise metrics across candidate qubit pairs. (2) \textbf{Optimization:} a constrained multi-objective solver selects $n$ couplers by minimizing a weighted cost that captures path length, error probability, congestion, overload, and sparsity. Hardware rules such as maximum degree and spatial separation are enforced as hard constraints. (3) \textbf{Outputs:} the algorithm returns the selected coupler set $\mathcal{L}$ along with metrics summarizing fidelity, path distribution, and structural balance across the modules.


This workflow enables principled inter-chip connectivity design: starting from raw hardware parameters, it produces a link set that is both feasible and near-optimal in terms of communication performance.

\subsection{\sol~ Optimization Workflow}

Building on the workflow in Section~\ref{sec:system_overview}, we design an algorithm that operationalizes coupler selection as a structured optimization procedure. The algorithm takes as input the calibrated hardware parameters (gate times, error rates, chip topology) and outputs a connectivity configuration that is both hardware-feasible and performance-aware.

To ensure physical feasibility, we enforce the following conditions:
\begin{align}
\deg(u) + x_{uv} &\le D_{\max}, 
\quad \forall\,u \in V_A,\\
\deg(v) + x_{uv} &\le D_{\max}, 
\quad \forall\,v \in V_B,\\
d_{\mathrm{phys}}\bigl((u_i, v_i), (u_j, v_j)\bigr) &> \delta,
\quad \forall\,i \neq j,\\
n &\le N_{\max}.
\end{align}

It proceeds in three phases:

\paragraph{1. Cost Matrix Construction.} 

For each candidate qubit pair $(i,j)$ across chips, we precompute a time-to-fidelity (TTF) cost that integrates latency and error rates into a single multiplicative factor. This results in an $n\times n$ cost matrix where each entry encodes the effective communication quality between a qubit pair. On-chip couplings are assigned baseline costs, while inter-chip candidates include both link latency and calibration-derived error penalties. Given: Quantum chip graphs $G_A = (V_A, E_A)$ and $G_B = (V_B, E_B)$. Physical error rates and fidelities for candidate couplers.
Degree limit $D_{\max}$, spacing constraint $\delta$, and number of required links $n$. Find:  a set of $n$ coupler links $\{(u_i, v_i)\}_{i=1}^n$ between $A$ and $B$ that minimizes the total cost in Equation~\ref{eq:cost_fn} while satisfying all constraints.

\paragraph{2. Iterative Link Selection.}
We cast coupler selection as a constrained multi-objective optimization. At each iteration, the solver evaluates candidate link sets according to the weighted cost function \ref{eq:cost_fn}, subject to architectural constraints such as maximum degree per node and physical spacing rules. Candidate sets are explored using a hybrid search strategy: greedy expansion for low-cost edges combined with local refinement (e.g., simulated annealing) to escape suboptimal configurations.

\paragraph{3. Validation and Output.}
Once a feasible set $\mathcal{L}$ of inter-chip couplers is selected, we validate the design by simulating routing load across benchmark circuits. The algorithm records on-/inter-chip path distributions, congestion hotspots, and predicted fidelity under the selected connectivity. These metrics are exported along with $\mathcal{L}$ for downstream compilation and benchmarking.

\medskip
In summary, the algorithm transforms raw hardware calibration data into actionable design choices by combining a fidelity-aware cost model with constrained optimization. This ensures that the resulting coupler set achieves a favorable trade-off between latency and fidelity while remaining robust across compiler backends and circuit families.



\section{Performance Evaluation}

In this section, we evaluate the performance of \sol~ with a focus on our cost model (e.g., Equation~\ref{eq:cost_fn}) with different coupler position selection. We focus on its relevance to On- and Inter-chip operations as well as fidelity.

\subsection{Implementation and Evaluation Settings}

\subsubsection{Implementation} \sol~ is implemented with Python 3.10 with Qiskit 1.2, and tested on a Google Cloud e2-highmem-16 instance with AMD Rome x86/64 processors. We benchmark a diverse set of circuits with the number of qubits (30 to 200). These include widely used quantum algorithms and subroutines.



\subsubsection{Modular and Chiplet system construction} 

We build our simulated modular and chiplet system with realistic superconducting quantum device models provided by IBM Quantum public backends~\cite{ibmFake_providerlatest}. These backends are built to mimic the behaviors of IBM Quantum systems using system snapshots. The system snapshots contain important information about the quantum system such as coupling map, basis gates, qubit properties, such as T1, T2, error rate, etc, which are useful for testing the transpiler and performing noisy simulations of the system.


To build an \sol~ system, we construct Inter-chip coupler links between multiple backends. We evaluated coupler-connected systems comprising 2 to 5 chips, varying from qubit counts and topologies, such as CairoV2 and Auckland with 27 qubits, Marrakesh with 156 qubits, to model both homogeneous configurations (identical chips) and heterogeneous ones (different chips combined). The resulting multi-chip systems spanned 54 to 312 qubits, allowing us to explore scaling behavior with chip count. By systematically varying coupler placements across these systems, we capture both the benefits and challenges of inter-chip connectivity in modular and chiplet architectures.


{\noindent \bf Parameter Settings:}  
In \sol~ cost model, e.g., Equation~\ref{eq:cost_fn}, tunable weights allow us to adjust the relative importance of each cost component when designing a coupler-aware chip-to-chip system. By varying these weights, the designer can prioritize different hardware considerations such as path fidelity, congestion, or structural balance. In our evaluation, we set $\alpha=\gamma=\delta=\epsilon=1$ and $\beta=10$, where the larger value of $\beta$ reflects the scaling of the effective path cost (Eq.~\ref{eq:effective_path_cost}) based on $T_1$ and $T_2$ times, measured in nanoseconds. As a result, the effective path cost values fall within $[0,1]$, while the other items span approximately $[1,10]$, ensuring that each term contributes comparably to the overall cost.

Additionally, following the IBM Flamingo technical report~\cite{ibmFlamingo2024}, we model CNOT gates on coupler links with a 3.5\% error rate over a 235\,ns operation, and approximate SWAP gates as incurring three times this error. By tuning these parameters, our framework supports both modular and chiplet systems.


\begin{figure*}
    \centering
    \includegraphics[width=0.95\linewidth]{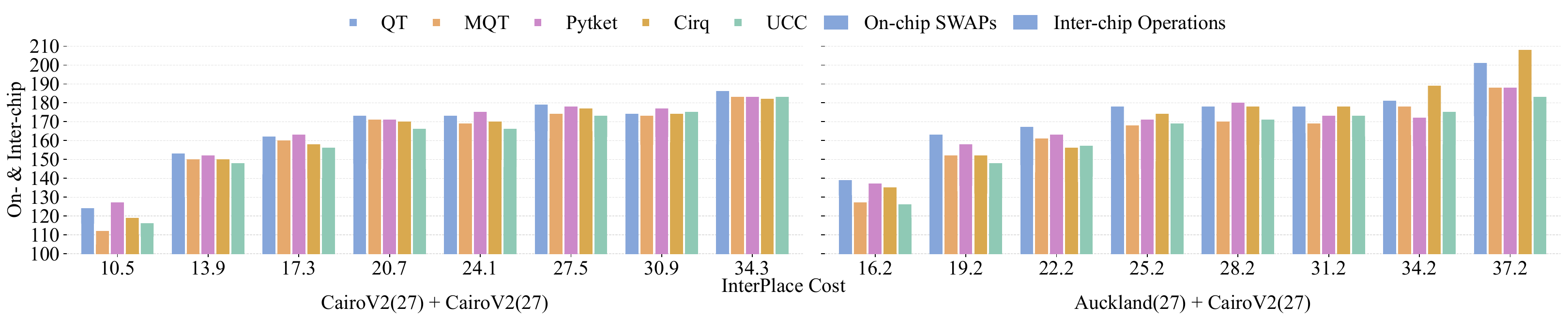}
    \caption{Relationship between \sol~ cost (Eq.~\ref{eq:cost_fn}) and Inter-chip operations + On-chip SWAPs for a random 40-qubit circuit of depth 10. Two configurations are shown: a homogeneous system of two identical CairoV2(27) chips (left) and a heterogeneous system composed of an Auckland(27) chip connected to a CairoV2(27) chip (right). Bars show On-chip (darker) and Inter-chip (lighter) SWAPs, with colors denoting compilers. Each group of bars corresponds to the same cost value at a specific coupler-link placement.}
    \label{fig:cost_vs_swaps}
\end{figure*}

\subsubsection{Compatibility and Baselines}
Notably, \sol~ optimizes to modular and chiplet quantum system at the manufacturing phase. This makes \sol~ fully compatible with existing quantum software ecosystem, such as quantum compilers and transpilation. These software-based processes aim to optimize qubit mapping, e.g., reducing number of SWAPs and improving algorithmic fidelity, based on specific hardware topologies. While \sol~ and existing compilers have similar objectives, they work at different layers. To fully understand the performance of \sol, we evaluate it with 5 popular quantum compilers: 
\begin{itemize}
    \item {\bf Qiskit-Transpiler}: IBM-Q’s default quantum compiler for monolithic devices~\cite{ibmTranspilerlatest}.
    \item {\bf MQT}: a framework for efficient machine-learning qubit mapping~\cite{mqtqmap} from Munich Quantum Toolkit; 
    \item {\bf Pytket}: a compiler and toolkit for quantum programming developed by Quantinuum~\cite{pytket2020}; 
    \item {\bf Cirq}: Google’s quantum circuit framework optimized for near-term quantum devices~\cite{cirq2023}; 
    \item {\bf UCC}: a lightweight and extensible compiler framework for quantum circuit transformation and optimization, developed as part of the Unitary Compiler Collection~\cite{ucc2025}.
\end{itemize}

As baseline comparisons, we emulate different Inter-chip linkage combinations and find (1) {\bf Lowest cost}: this is the solution \sol~ provides. (2) {\bf Median cost}: it is one representative random selection. (3) {\bf Highest cost}: this is the linkage that offers the highest cost value based on our model.

\subsubsection{Evaluation Metrics}: 
\sol~ aims to provide flexibility to quantum compilers to reduce the costly operations and improve fidelity.  We focus on the following metrics from the transpiled circuits, e.g., circuit layouts after compilation.

\begin{itemize}
    \item \textbf{Inter-chip Operations:} Total two-qubit gates executed across chips along selected couplers. It represents a type of most expensive operations in a modular and chiplet quantum system due to higher errors and longer gate times. 

    \item \textbf{On-chip SWAPs:}  Total number of SWAP gates executed within individual chips, e.g., without using Inter-chip coupler links. This is a commonly used metric for compilers in single-chip superconducting platforms.

    \item \textbf{Fidelity:} It quantifies how closely the transpiled circuit matches the ideal output by compounding error from two-qubit gates, SWAPs (treated as three CNOTs), and Inter-chip couplers.

    \item \textbf{\sol~ Cost:} the weighted objective in Equation~\eqref{eq:cost_fn}, combining Average Path Length(APL), Effective Path Cost(EPC), Congestion(Cong), Qubit Overload(Over), and Sparsity(Spar) to rank link sets.
    
\end{itemize}

\subsubsection{Workloads}: Our workload include both random circuits and algorithmic circuits to fully evaluate \sol.  

We utilize random circuits to benchmark our systems and study the effectiveness of \sol. Those circuits generated using Qiskit's built-in utilities, commonly used for general-purpose benchmarking.

To understand how \sol~ affects algorithmic fidelity, we evaluate it with following circuits:
\textbf{CAT}: Prepares Schr{\"o}-dinger cat states ~\cite{mirrahimi2014cat}. 
\textbf{QAOA} (Quantum Approximate Optimization Algorithm): A hybrid quantum-classical variational algorithm for solving combinatorial optimization problems~\cite{farhi2014qaoa}. 
\textbf{GHZ}: Constructs Greenberger-Horne-Zeilinger (GHZ) states for nonlocality and entanglement verification~\cite{greenberger1989going}. 
\textbf{QFT}: Quantum Fourier Transform, fundamental quantum subroutine underlying many quantum algorithms, including Shor’s factoring algorithm~\cite{nielsen2010quantum}.

\subsection{Effectiveness of \sol~ Cost Modeling}

\begin{figure*}[t]
    \centering
    \includegraphics[width=0.95\linewidth]{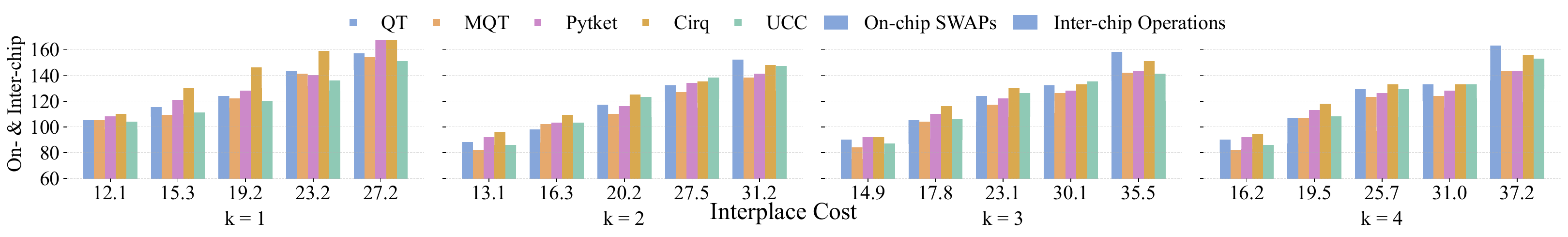}
    \caption{SWAP counts versus cost for coupler link sets of different sizes. Each panel shows results for a random 30-qubit circuit with $k=1$--$4$ couplers placed in a system with one Auckland(27) and one CairoV2(27). Bars show SWAP counts, separated into On-chip (dark) and Inter-chip (light) components, while the line marks the total SWAP count. Colors denote five compilers. Increasing $k$ enlarges the design space for coupler placement, providing more flexibility for routing.}
    \label{fig:k_links}
\end{figure*}

\begin{table*}[h]
\centering
\caption{Inter-Chip Operations under Different \sol~ Cost Values in a 3- to 5-Chips System with 5 compilers}
\label{tab:2}
\resizebox{0.95\textwidth}{!}{%
\begin{tabular}{|c|ccc|c|
ccc|ccc|ccc|ccc|ccc|}
\hline
\multirow{2}{*}{\textbf{Chips}}
& \multicolumn{3}{c|}{\textbf{\sol~ Cost}}
& \multirow{2}{*}{\textbf{Circuit}}
& \multicolumn{3}{c|}{\textbf{QT}} 
& \multicolumn{3}{c|}{\textbf{MQT}} 
& \multicolumn{3}{c|}{\textbf{Pytket}} 
& \multicolumn{3}{c|}{\textbf{Cirq}} 
& \multicolumn{3}{c|}{\textbf{UCC}} 
\\ \cline{2-4}\cline{6-20}
& Lowest & Median & Highest 
& 
& L & M & H
& L & M & H
& L & M & H
& L & M & H
& L & M & H \\
\hline

\multirow{2}{*}{\textbf{3}} 
& \multirow{2}{*}{18.37}
& \multirow{2}{*}{28.49}
& \multirow{2}{*}{36.77}
& Random(60)  & 67 & 74 & 82
         & 54 & 70 & 78 
         & 78 & 96 & 105 
         & 84 & 98 & 112 
         & 70 & 82 & 88 \\

& &&& Random(80) & 152  & 178 & 195 
         & 150 & 186 & 202
         & 154 & 198 & 219 
         & 157 & 192 & 223 
        & 148 & 189 & 202 \\
\hline

\multirow{2}{*}{\textbf{4}} 
& \multirow{2}{*}{20.86}
& \multirow{2}{*}{38.24}
& \multirow{2}{*}{46.73}
& Random(80)   & 124  & 148 & 166 
         & 129 & 156 & 172 
         & 135 & 188 & 193 
         & 157 & 172 & 198 
        & 118 & 126 & 177 \\

& &&& Random(100) &  227  & 243 & 268 
         & 201 & 232 & 243 
         & 212 & 244 & 263 
         & 234 & 259 & 278 
         & 207 & 215 & 258 \\

\hline

\multirow{2}{*}{\textbf{5}} 
& \multirow{2}{*}{26.67}
& \multirow{2}{*}{47.63}
& \multirow{2}{*}{56.19}
& Random(100)  &  196  & 216 & 222 
         & 202 & 232 & 248 
         & 212 & 257 & 283 
         & 234 & 278 & 298 
         & 198 & 215 & 237 \\

& &&& Random(130) & 366 & 421 & 445
        & 372 & 419 & 451
        & 378 & 427 & 457
        & 394 & 435 & 452
        & 402 & 443 & 463 \\

\hline

\end{tabular}
}
\end{table*}

In this subsection, we aim to demonstrate the effectiveness of \sol~ cost modeling and its relevance to On-chip SWAPs and Inter-chip Operations. 

\subsubsection{4-Link 2-Chip Systems}
Firstly, we build two systems to evaluate \sol's performance, a homogeneous system with two identical CairoV2(27) chips, e.g., the same error models and topologies, and a heterogeneous system with one Auckland(27) and one CairoV2(27). Chips in these systems are connected with 4 Inter-chip coupler links, e.g., same as IBM's L couplers. We conduct these experiments with a 40-qubit random circuit with depth 10.

Figure~\ref{fig:cost_vs_swaps} reports the combined On-Chip SWAP counts and Inter-Chip operations across 5 compilers in both homogeneous and heterogeneous systems. 
Obviously, we clearly discover a trend that over \sol~ cost results in lower number of total Inter-chip operations and On-chip SWAPs. 




In the homogeneous case, the combined Inter-chip operations and On-chip SWAPs decrease markedly as the cost metric improves. Qiskit compiler (QT) decreases from $182$ total, i.e., 27 Inter-chip operations and 155 On-chip SWAPs, at the highest cost ($34.3$) to $119$ total, i.e., 9 Inter-chip operations and 110 On-chip SWAPs, at the lowest cost ($10.5$), a reduction of roughly $63$ or 34.6\%. MQT reduces 71 total or $38.8\%$ from 183 total, i.e., 22 Inter-chip operations, to 112 total, i.e., 7 Inter-chip operations.

The heterogeneous case represents a more challeng setting due to different error models and topologies from two chips. We find a similar pattern, with counts falling from $208$ at cost $37.2$ to $135$ at a cost $16.2$, corresponding to a decrease of $73$ or 35.1\%, at Qiskit Compiler. UCC reduces 57 total or 31.1\%, and 16 Inter-chip operations from 23 to 7, in the lowest and highest cost, respectively. These consistent reductions confirm that lower-cost link placement yield lower On-chip SWAPs and Inter-chip operations overhead across both architectures and among all 5 tested compilers.


Results demonstrate the central contribution of our cost model. By integrating path length, congestion, and noise awareness into unified formulation, the model provides a reliable predictor of Inter-chip operations and On-chip SWAPs. 


\subsubsection{Different Numbers of Inter-Chip Links Impact}

Next, we study how the number of Inter-chip links can affect \sol's performance. 

Based on a system with one Auckland (27) and one CairoV2 (27), 30-qubit 10-depth random circuit, Figure~\ref{fig:k_links} compares Inter-chip operations and On-chip SWAP across 5 compilers when varying the number of available Inter-chip couplers from $k=1$ to $k=4$. This experiment is critical because increasing the number of available links enlarges compiler routing freedom, but it also changes the optimization landscape. Evaluating different $k$ values shows how \sol~ cost model generalizes the size of coupler links across system sizes and compilers.





We clearly discover the same trend from 4 experiments such that a lower \sol~ cost leads to lower combined Inter-chip operations and On-chip SWAPs. 
Specifically, with only $k=1$ link, the combination of Inter-chip operations and On-chip SWAPs is tightly constrained (e.g., Cirq ranges from $110$ to $167$), since limited routing forces long paths and leaves little room to avoid congestion. Increasing to $k=2$ expands variability: MQT ranges from $82$ at cost $13.1$ to $143$ at cost $31.2$, showing that while extra links enable better placement, poor choices still inflate Inter-chip traffic. At $k=3$, spreads widen further, with Pytket ranging $92$–$143$ depending on cost, highlighting the growing impact of link selection. With $k=4$, our cost model is most effective: Qiskit drops from $163$ at cost $37.2$ to $90$ at cost $16.2$, and differences across \sol~ cost exceed $50$. These results confirm that careful, cost-guided link placement is crucial for reducing congestion and Inter-chip overhead.

These results highlight two key insights. First, increasing $k$ does not guarantee better performance. Instead, it expands the design space, making the placement of Inter-chip links even more critical. This explains why the best achievable cost value also increases with larger $k$: additional links create more routing possibilities, but not all of them lead to efficient mappings. Second, our cost-driven framework consistently identifies high-quality link placements, enabling compilers to leverage the extra routing freedom of larger $k$ without introducing congestion or error hot-spots. This suggests a promising direction for hardware with tunable link budgets, where compiler-guided cost metrics ensure that additional couplers translate into real fidelity and SWAP improvements rather than wasted overhead.

\subsection{Scalability of \sol~ Cost Modeling}
\label{sec:s}
In this subsection, we investigate the scalability of \sol. We study two axes: (i) the number of chips to be connected by Inter-chip couplers and (ii) the chip size, e.g., qubits per chip. \sol~ selects a given number of Inter-chip links by minimizing the cost. 




\subsubsection{Different Chip Counts} We scale the system to \(m\!\in\!\{3,4,5\}\) chips, i.e., Auckland(27), under a fixed per-pair link budget \(k\) (e.g., \(k{=}4\)). As \(m\) increases, \sol~ consistently lowers total cost and substantially reduces Inter-chip operations across all compilers. 
Table~\ref{tab:2} presents the results with random circuits. As we can see, \sol~ cuts the cost metric by \(50.0\%\) for \(m{=}3\) (36.77\(\rightarrow\)18.37), \(55.4\%\) for \(m{=}4\) (46.73\(\rightarrow\)20.86), and \(52.5\%\) for \(m{=}5\) (56.19\(\rightarrow\)26.67).

These reductions translate directly into fewer Inter-chip operations across compilers. For example, with 4 chips, UCC drops from 177 to 118 (33.3\% savings), while at 5 chips Pytket falls from 283 to 212 (25.1\%). Improvements of this scale are consistent across QT, MQT, Pytket, Cirq, and UCC, confirming that cost-aware link placement is increasingly critical as modular systems scale. Overall, \sol\ demonstrates strong scalability by mitigating Inter-chip overhead while preserving compiler flexibility in larger multi-chip systems.

\begin{figure}
    \centering
    \includegraphics[width=0.85\linewidth]{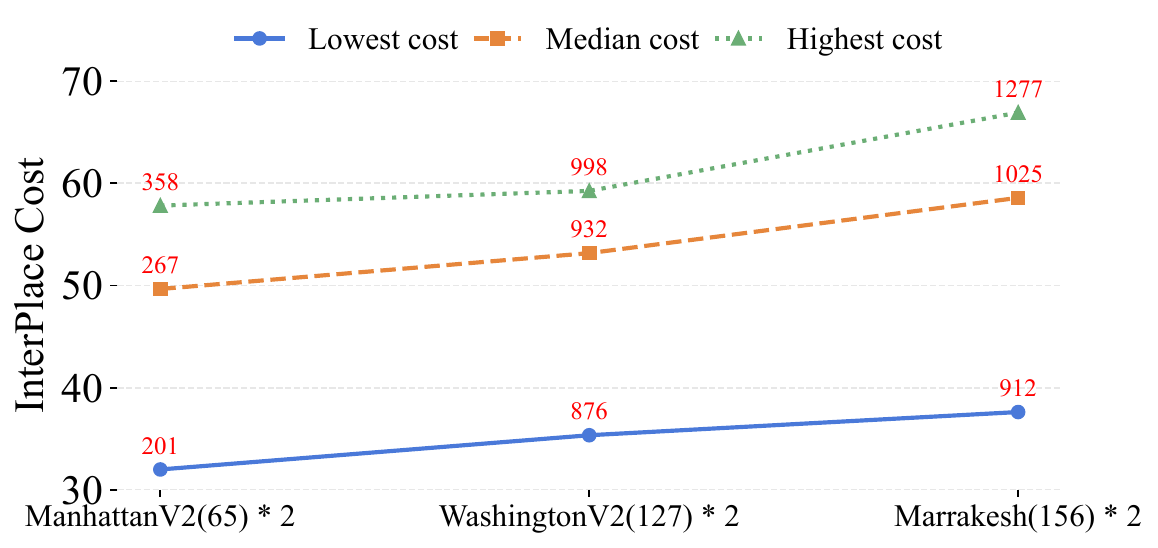}
    \caption{Cost comparison of the lowest (\sol), median, and highest-cost coupler link strategies across different chip sizes. Red numbers indicate the Inter-chip operations under the UCC compiler for random circuits with 100, 150, and 200 qubits on two-chip systems of 130, 254, and 312 qubits.}
    \label{fig:Lchip}
\end{figure}

\subsubsection{Different Chip Sizes}

Next, we scale the system size from \(n{=}27\) to \(n\!\in\!\{65, 127, 156\}\), corresponding to ManhattanV2(65), WashingtonV2(127), and Marrakesh(156), while keeping the number of chips fixed at two. For a 4-link Inter-chip configuration, the routed SWAP count increases roughly linearly with \(n\), reflecting the longer detours required in heavy-hex topologies. By anchoring links at low-cost positions, our solver effectively curbs this growth.


Figure~\ref{fig:Lchip} visually compares the cost distributions for the lowest, median, and highest strategies across different chip sizes. The \sol~ approach consistently yields the lowest cost, achieving reductions of 44.7\%, 40.3\%, and 43.8\%. Specifically, costs decrease from 57.82\(\rightarrow\)32.00, 59.26\(\rightarrow\)35.35, and 66.92\(\rightarrow\)37.62 for the three system sizes, respectively. The Inter-chip operations for Random(200) with depth 10, are 912, 1025, and 1277 operations under the lowest-, median-, and highest-cost placements. 

These results confirm that the cost advantage of \sol~ is preserved across different system scales, underscoring its scalability. While costs naturally rise with system size, the growth rate is significantly lower when guided by \sol. This performance gain stems from its optimized coupler selection and efficient resource allocation strategy.



\subsection{Fidelity under Different \sol~ Costs}
Finally, we study whether \sol~ cost can translate to improved algorithmic fidelity. 
Due to the limitations of classical simulations, we only present the results from circuits with up to 30 qubits.

\begin{figure}[]
    \centering
    \includegraphics[width=1\linewidth]{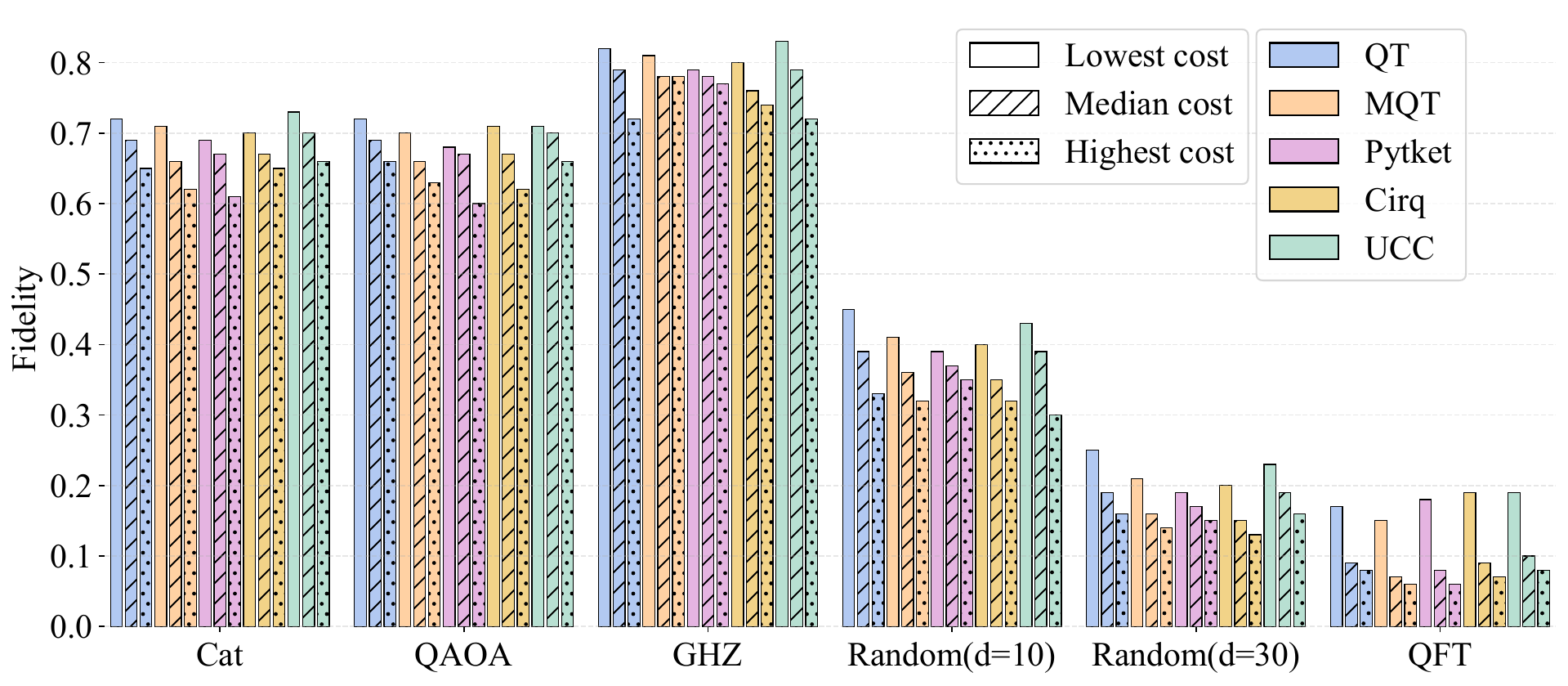}
    \caption{Fidelity comparison of our optimized coupler selection (\sol) versus a random baseline across six 30-qubit circuit families compiled with five backends on a system with two connected Auckland(27). Each category shows ten bars: the first five are \sol\ results, one per compiler in fixed order; the last five are the matched baseline.}
    \label{fig:Fidelity}
\end{figure}

In Fig.~\ref{fig:Fidelity}, we compare \sol~ with the lowest cost corresponding to the median and highest cost across six 30-qubit circuits and 5 compilers, evaluated on a system composed of two 27-qubit Auckland chips. Each compiler group reports fidelity under 3 cost regimes, showing how the Inter-chip coupler operations directly fidelity.



Obviously, \sol~ consistently outperforms because the link set is chosen by minimizing a composite cost objective that balances path length, error rate, congestion, overload, and sparsity. The optimized configuration achieves a total cost of 16.96, compared with the median case at 33.18 and the highest case at 37.69. 

In terms of fidelity, the improvements are substantial across all benchmarks. For example, in the Random circuit with depth $30$, fidelity increases from $19\%$ (highest-cost) to $25\%$ (lowest-cost), a relative gain of $31.6\%$. The QFT circuit exhibits the most dramatic effect: fidelity improves from only $8\%$ (highest-cost) to $17\%$ (lowest-cost) under \sol, $53.0\%$ increased. These gains correlate directly with reduced SWAP overhead; for instance, in QFT compiled by MQT, the SWAP count decreases from $256$ (highest-cost) to $195$ under \sol, while Inter-chip operations drop from $34$ to $22$.

\section{Conclusion}

We introduced \sol, a hardware-aware framework for optimizing inter-chip coupler placement in modular and chiplet-based quantum systems. In contrast to prior approaches that adapt compilers to fixed hardware, \sol~integrates physical constraints with logical communication objectives to co-design coupler locations. Its unified cost model incorporates path length, latency–fidelity trade-offs, congestion, overload, and sparsity, all evaluated prior to fabrication.

The framework requires time to search for optimal coupler placements, but this process occurs entirely in the preprocessing stage. Exploring candidate inter-chip connections and compiler behaviors can take several hours per system size; however, this search is performed only once per hardware design point, before fabrication. The cost is therefore amortized across all future workloads and compilers, ensuring that manufactured systems achieve lower SWAP counts and higher fidelity. Modest preprocessing time thus enables long-term scalability and robustness for modular quantum architectures.

Our evaluation on both homogeneous and heterogeneous multi-chip systems shows that \sol~effectively reduces SWAPs and inter-chip operations while improving fidelity across different compilers. The framework also scales with chip count and system size, achieving cost reductions of up to 55.4\% and fidelity improvements of up to 53.0\% compared to baseline placements. Although optimization introduces offline computation time, this overhead is absorbed at design time and delivers lasting performance benefits.

In summary, \sol~bridges physical feasibility and logical efficiency in modular and chiplet system architectures, advancing the path toward scalable, high-fidelity quantum processors.

\section{Acknowledgment}
This research was supported in part by the National Science Foundation (NSF) under grant agreements 2301884, 2329020, 2335788, and 2343535.

\bibliographystyle{ACM-Reference-Format}  
\bibliography{refs}


\begin{thebibliography}{53}


\ifx \showCODEN    \undefined \def \showCODEN     #1{\unskip}     \fi
\ifx \showDOI      \undefined \def \showDOI       #1{#1}\fi
\ifx \showISBNx    \undefined \def \showISBNx     #1{\unskip}     \fi
\ifx \showISBNxiii \undefined \def \showISBNxiii  #1{\unskip}     \fi
\ifx \showISSN     \undefined \def \showISSN      #1{\unskip}     \fi
\ifx \showLCCN     \undefined \def \showLCCN      #1{\unskip}     \fi
\ifx \shownote     \undefined \def \shownote      #1{#1}          \fi
\ifx \showarticletitle \undefined \def \showarticletitle #1{#1}   \fi
\ifx \showURL      \undefined \def \showURL       {\relax}        \fi
\providecommand\bibfield[2]{#2}
\providecommand\bibinfo[2]{#2}
\providecommand\natexlab[1]{#1}
\providecommand\showeprint[2][]{arXiv:#2}

\bibitem[ank({[n.\,d.]})]%
        {ankaa}
 \bibinfo{year}{[n.\,d.]}\natexlab{}.
\newblock \bibinfo{howpublished}{\url{https://investors.rigetti.com/static-files/fbac3801-223f-4f0f-a207-47d25084a1d7}}.
\newblock
\newblock
\shownote{[Accessed 20-08-2025]}.


\bibitem[pyt({[n.\,d.]})]%
        {pytket2020}
 \bibinfo{year}{[n.\,d.]}\natexlab{}.
\newblock \bibinfo{title}{{G}it{H}ub - {C}{Q}{C}{L}/tket: {S}ource code for the {T}{K}{E}{T} quantum compiler, {P}ython bindings and utilities --- github.com}.
\newblock \bibinfo{howpublished}{\url{https://github.com/CQCL/tket}}.
\newblock
\newblock
\shownote{[Accessed 01-09-2025]}.


\bibitem[ucc({[n.\,d.]})]%
        {ucc2025}
 \bibinfo{year}{[n.\,d.]}\natexlab{}.
\newblock \bibinfo{title}{{G}it{H}ub - unitaryfoundation/ucc: {U}nitary {C}ompiler {C}ollection --- github.com}.
\newblock \bibinfo{howpublished}{\url{https://github.com/unitaryfoundation/ucc}}.
\newblock
\newblock
\shownote{[Accessed 01-09-2025]}.


\bibitem[iyq({[n.\,d.]})]%
        {iyq}
 \bibinfo{year}{[n.\,d.]}\natexlab{}.
\newblock \bibinfo{title}{International Year of Quantum Sci@ ence and Technology}.
\newblock \bibinfo{howpublished}{\url{https://www.quantum2025.org/}}.
\newblock
\newblock
\shownote{Accessed: 2025-04-09}.


\bibitem[blo({[n.\,d.]})]%
        {blogMeetWillow}
 \bibinfo{year}{[n.\,d.]}\natexlab{}.
\newblock \bibinfo{title}{{M}eet {W}illow, our state-of-the-art quantum chip --- blog.google}.
\newblock \bibinfo{howpublished}{\url{https://blog.google/technology/research/google-willow-quantum-chip/}}.
\newblock
\newblock
\shownote{[Accessed 21-08-2025]}.


\bibitem[ibm({[n.\,d.]})]%
        {ibmTranspilerlatest}
 \bibinfo{year}{[n.\,d.]}\natexlab{}.
\newblock \bibinfo{title}{transpiler (latest version) | {I}{B}{M} {Q}uantum {D}ocumentation --- quantum.cloud.ibm.com}.
\newblock \bibinfo{howpublished}{\url{https://quantum.cloud.ibm.com/docs/en/api/qiskit/transpiler}}.
\newblock
\newblock
\shownote{[Accessed 20-08-2025]}.


\bibitem[ibm(2025)]%
        {ibmFake_providerlatest}
 \bibinfo{year}{2025}\natexlab{}.
\newblock \bibinfo{title}{fake\_provider (latest version) | {I}{B}{M} {Q}uantum {D}ocumentation --- quantum.cloud.ibm.com}.
\newblock \bibinfo{howpublished}{\url{https://quantum.cloud.ibm.com/docs/en/api/qiskit-ibm-runtime/fake-provider}}.
\newblock
\newblock
\shownote{[Accessed 21-08-2025]}.


\bibitem[Abdikhakimov(2024)]%
        {abdikhakimov2024preparing}
\bibfield{author}{\bibinfo{person}{Islombek Abdikhakimov}.} \bibinfo{year}{2024}\natexlab{}.
\newblock \showarticletitle{Preparing for a Quantum Future: Strategies for Strengthening International Data Privacy in the Face of Evolving Technologies}.
\newblock \bibinfo{journal}{\emph{International Journal of Law and Policy}} \bibinfo{volume}{2}, \bibinfo{number}{5} (\bibinfo{year}{2024}), \bibinfo{pages}{42--46}.
\newblock


\bibitem[Arute et~al\mbox{.}(2019)]%
        {arute2019quantum}
\bibfield{author}{\bibinfo{person}{Frank Arute}, \bibinfo{person}{Kunal Arya}, \bibinfo{person}{Ryan Babbush}, {et~al\mbox{.}}} \bibinfo{year}{2019}\natexlab{}.
\newblock \showarticletitle{Quantum supremacy using a programmable superconducting processor}.
\newblock \bibinfo{journal}{\emph{Nature}} \bibinfo{volume}{574}, \bibinfo{number}{7779} (\bibinfo{year}{2019}), \bibinfo{pages}{505--510}.
\newblock


\bibitem[Bandic et~al\mbox{.}(2023)]%
        {Bandic2023}
\bibfield{author}{\bibinfo{person}{Zoran Bandic} {et~al\mbox{.}}} \bibinfo{year}{2023}\natexlab{}.
\newblock \showarticletitle{Mapping Quantum Circuits to Modular Architectures with QUBO}.
\newblock \bibinfo{journal}{\emph{arXiv preprint arXiv:2305.06687}} (\bibinfo{date}{May} \bibinfo{year}{2023}).
\newblock
\urldef\tempurl%
\url{https://arxiv.org/abs/2305.06687}
\showURL{%
\tempurl}


\bibitem[Barends et~al\mbox{.}(2014)]%
        {barends2014superconducting}
\bibfield{author}{\bibinfo{person}{Rami Barends}, \bibinfo{person}{Julian Kelly}, \bibinfo{person}{Anthony Megrant}, \bibinfo{person}{Daniel Sank}, \bibinfo{person}{Evan Jeffrey}, \bibinfo{person}{Yu Chen}, \bibinfo{person}{Yi Yin}, \bibinfo{person}{Ben Chiaro}, \bibinfo{person}{Josh Mutus}, \bibinfo{person}{Charles Neill}, {et~al\mbox{.}}} \bibinfo{year}{2014}\natexlab{}.
\newblock \showarticletitle{Superconducting quantum circuits at the surface code threshold for fault tolerance}.
\newblock \bibinfo{journal}{\emph{Nature}} \bibinfo{volume}{508}, \bibinfo{number}{7497} (\bibinfo{year}{2014}), \bibinfo{pages}{500--503}.
\newblock
\urldef\tempurl%
\url{https://doi.org/10.1038/nature13171}
\showDOI{\tempurl}


\bibitem[Baseri et~al\mbox{.}(2024)]%
        {baseri2024navigating}
\bibfield{author}{\bibinfo{person}{Yaser Baseri}, \bibinfo{person}{Vikas Chouhan}, {and} \bibinfo{person}{Abdelhakim Hafid}.} \bibinfo{year}{2024}\natexlab{}.
\newblock \showarticletitle{Navigating quantum security risks in networked environments: A comprehensive study of quantum-safe network protocols}.
\newblock \bibinfo{journal}{\emph{Computers \& Security}}  \bibinfo{volume}{142} (\bibinfo{year}{2024}), \bibinfo{pages}{103883}.
\newblock


\bibitem[Cao et~al\mbox{.}(2019)]%
        {cao2019quantum}
\bibfield{author}{\bibinfo{person}{Yudong Cao}, \bibinfo{person}{Jonathan Romero}, \bibinfo{person}{Jonathan~P Olson}, \bibinfo{person}{Matthias Degroote}, \bibinfo{person}{Peter~D Johnson}, \bibinfo{person}{M{\'a}ria Kieferov{\'a}}, \bibinfo{person}{Ian~D Kivlichan}, \bibinfo{person}{Tim Menke}, \bibinfo{person}{Borja Peropadre}, \bibinfo{person}{Nicolas P~D Sawaya}, \bibinfo{person}{Sukin Sim}, \bibinfo{person}{Libor Veis}, {and} \bibinfo{person}{Al{\'a}n Aspuru-Guzik}.} \bibinfo{year}{2019}\natexlab{}.
\newblock \showarticletitle{Quantum Chemistry in the Age of Quantum Computing}.
\newblock \bibinfo{journal}{\emph{Chemical Reviews}} \bibinfo{volume}{119}, \bibinfo{number}{19} (\bibinfo{year}{2019}), \bibinfo{pages}{10856--10915}.
\newblock


\bibitem[Dai et~al\mbox{.}(2025)]%
        {dai2025quantum}
\bibfield{author}{\bibinfo{person}{Shuhong Dai}, \bibinfo{person}{Nishant Saurabh}, \bibinfo{person}{Qingle Wang}, \bibinfo{person}{Jiawei Nian}, \bibinfo{person}{Shuwen Kan}, \bibinfo{person}{Ying Mao}, {and} \bibinfo{person}{Long Cheng}.} \bibinfo{year}{2025}\natexlab{}.
\newblock \showarticletitle{Quantum Reinforcement Learning for QoS-Aware Real-Time Job Scheduling in Cloud Systems}.
\newblock \bibinfo{journal}{\emph{IEEE Systems Journal}} (\bibinfo{year}{2025}).
\newblock


\bibitem[Du et~al\mbox{.}(2025)]%
        {du2025hardware}
\bibfield{author}{\bibinfo{person}{Zefan Du}, \bibinfo{person}{Shuwen Kan}, \bibinfo{person}{Samuel Stein}, \bibinfo{person}{Zhiding Liang}, \bibinfo{person}{Ang Li}, {and} \bibinfo{person}{Ying Mao}.} \bibinfo{year}{2025}\natexlab{}.
\newblock \showarticletitle{Hardware-aware Compilation for Chip-to-Chip Coupler-Connected Modular Quantum Systems}.
\newblock \bibinfo{journal}{\emph{arXiv preprint arXiv:2505.09036}} (\bibinfo{year}{2025}).
\newblock


\bibitem[Du et~al\mbox{.}(2024a)]%
        {du2024hardware}
\bibfield{author}{\bibinfo{person}{Zefan Du}, \bibinfo{person}{Yanni Li}, \bibinfo{person}{Zijian Mo}, \bibinfo{person}{Wenqi Wei}, \bibinfo{person}{Juntao Chen}, \bibinfo{person}{Rajkumar Buyya}, {and} \bibinfo{person}{Ying Mao}.} \bibinfo{year}{2024}\natexlab{a}.
\newblock \showarticletitle{Hardware-aware circuit cutting and distributed qubit mapping for connected quantum systems}.
\newblock \bibinfo{journal}{\emph{arXiv preprint arXiv:2412.18458}} (\bibinfo{year}{2024}).
\newblock


\bibitem[Du et~al\mbox{.}(2024b)]%
        {du2024efficient}
\bibfield{author}{\bibinfo{person}{Zefan Du}, \bibinfo{person}{Wenrui Zhang}, \bibinfo{person}{Wenqi Wei}, \bibinfo{person}{Juntao Chen}, \bibinfo{person}{Tao Han}, \bibinfo{person}{Zhiding Liang}, {and} \bibinfo{person}{Ying Mao}.} \bibinfo{year}{2024}\natexlab{b}.
\newblock \showarticletitle{Efficient Circuit Cutting and Scheduling in a Multi-Node Quantum System with Dynamic EPR Pairs}.
\newblock \bibinfo{journal}{\emph{arXiv preprint arXiv:2412.18709}} (\bibinfo{year}{2024}).
\newblock


\bibitem[D’Onofrio et~al\mbox{.}(2023)]%
        {d2023distributed}
\bibfield{author}{\bibinfo{person}{Anthony D’Onofrio}, \bibinfo{person}{Amir Hossain}, \bibinfo{person}{Lesther Santana}, \bibinfo{person}{Naseem Machlovi}, \bibinfo{person}{Samuel Stein}, \bibinfo{person}{Jinwei Liu}, \bibinfo{person}{Ang Li}, {and} \bibinfo{person}{Ying Mao}.} \bibinfo{year}{2023}\natexlab{}.
\newblock \showarticletitle{Distributed quantum learning with co-management in a multi-tenant quantum system}. In \bibinfo{booktitle}{\emph{2023 IEEE International Conference on Big Data (BigData)}}. IEEE, \bibinfo{pages}{221--228}.
\newblock


\bibitem[Elsharkawy et~al\mbox{.}(2025)]%
        {elsharkawy2025integration}
\bibfield{author}{\bibinfo{person}{Amr Elsharkawy}, \bibinfo{person}{Xiao-Ting~Michelle To}, \bibinfo{person}{Philipp Seitz}, \bibinfo{person}{Yanbin Chen}, \bibinfo{person}{Yannick Stade}, \bibinfo{person}{Manuel Geiger}, \bibinfo{person}{Qunsheng Huang}, \bibinfo{person}{Xiaorang Guo}, \bibinfo{person}{Muhammad~Arslan Ansari}, \bibinfo{person}{Christian~B Mendl}, {et~al\mbox{.}}} \bibinfo{year}{2025}\natexlab{}.
\newblock \showarticletitle{Integration of quantum accelerators with high performance computing—a review of quantum programming tools}.
\newblock \bibinfo{journal}{\emph{ACM Transactions on Quantum Computing}} \bibinfo{volume}{6}, \bibinfo{number}{3} (\bibinfo{year}{2025}), \bibinfo{pages}{1--46}.
\newblock


\bibitem[Farhi et~al\mbox{.}(2014)]%
        {farhi2014qaoa}
\bibfield{author}{\bibinfo{person}{Edward Farhi}, \bibinfo{person}{Jeffrey Goldstone}, {and} \bibinfo{person}{Sam Gutmann}.} \bibinfo{year}{2014}\natexlab{}.
\newblock \showarticletitle{A quantum approximate optimization algorithm}.
\newblock \bibinfo{journal}{\emph{arXiv preprint arXiv:1411.4028}} (\bibinfo{year}{2014}).
\newblock


\bibitem[Gambetta and Mandelbaum(2024)]%
        {ibmFlamingo2024}
\bibfield{author}{\bibinfo{person}{Jay Gambetta} {and} \bibinfo{person}{Ryan Mandelbaum}.} \bibinfo{year}{2024}\natexlab{}.
\newblock \bibinfo{title}{{IBM Quantum delivers on performance challenge made two years ago}}.
\newblock \bibinfo{howpublished}{\url{https://www.ibm.com/quantum/blog/qdc-2024}}.
\newblock
\newblock
\shownote{Accessed: 2025-04-09}.


\bibitem[Gambetta et~al\mbox{.}(2017)]%
        {gambetta2017building}
\bibfield{author}{\bibinfo{person}{Jay~M Gambetta}, \bibinfo{person}{Jerry~M Chow}, {and} \bibinfo{person}{Matthias Steffen}.} \bibinfo{year}{2017}\natexlab{}.
\newblock \showarticletitle{Building logical qubits in a superconducting quantum computing system}.
\newblock \bibinfo{journal}{\emph{npj Quantum Information}} \bibinfo{volume}{3}, \bibinfo{number}{1} (\bibinfo{year}{2017}), \bibinfo{pages}{2}.
\newblock


\bibitem[{Google Quantum AI}(2023)]%
        {cirq2023}
\bibfield{author}{\bibinfo{person}{{Google Quantum AI}}.} \bibinfo{year}{2023}\natexlab{}.
\newblock \bibinfo{title}{Cirq}.
\newblock \bibinfo{howpublished}{\url{https://quantumai.google/cirq}}.
\newblock
\newblock
\shownote{Version 1.2.0}.


\bibitem[Greenberger et~al\mbox{.}(2007)]%
        {greenberger1989going}
\bibfield{author}{\bibinfo{person}{Daniel~M. Greenberger}, \bibinfo{person}{Michael~A. Horne}, {and} \bibinfo{person}{Anton Zeilinger}.} \bibinfo{year}{2007}\natexlab{}.
\newblock \bibinfo{title}{Going Beyond Bell's Theorem}.
\newblock
\newblock
\showeprint[arxiv]{0712.0921}~[quant-ph]
\urldef\tempurl%
\url{https://arxiv.org/abs/0712.0921}
\showURL{%
\tempurl}


\bibitem[Jeng et~al\mbox{.}(2025)]%
        {Jeng2025}
\bibfield{author}{\bibinfo{person}{Mingyoung~Jessica Jeng} {et~al\mbox{.}}} \bibinfo{year}{2025}\natexlab{}.
\newblock \showarticletitle{Modular Compilation for Quantum Chiplet Architectures}.
\newblock \bibinfo{journal}{\emph{arXiv preprint arXiv:2501.08478}} (\bibinfo{date}{January} \bibinfo{year}{2025}).
\newblock
\urldef\tempurl%
\url{https://arxiv.org/abs/2501.08478}
\showURL{%
\tempurl}


\bibitem[Kan et~al\mbox{.}(2024)]%
        {kan2024scalable}
\bibfield{author}{\bibinfo{person}{Shuwen Kan}, \bibinfo{person}{Zefan Du}, \bibinfo{person}{Miguel Palma}, \bibinfo{person}{Samuel~A. Stein}, \bibinfo{person}{Chenxu Liu}, \bibinfo{person}{Wenqi Wei}, \bibinfo{person}{Juntao Chen}, \bibinfo{person}{Ang Li}, {and} \bibinfo{person}{Ying Mao}.} \bibinfo{year}{2024}\natexlab{}.
\newblock \showarticletitle{Scalable Circuit Cutting and Scheduling in a Resource-constrained and Distributed Quantum System}. In \bibinfo{booktitle}{\emph{2024 IEEE Quantum Week Conference (QCE)}} (Montréal, QC, Canada). \bibinfo{publisher}{IEEE}.
\newblock


\bibitem[Kjaergaard et~al\mbox{.}(2020)]%
        {kjaergaard2020superconducting}
\bibfield{author}{\bibinfo{person}{Morten Kjaergaard}, \bibinfo{person}{Mollie~E Schwartz}, \bibinfo{person}{Jochen Braum{\"u}ller}, \bibinfo{person}{Philip Krantz}, \bibinfo{person}{Joel I-J Wang}, \bibinfo{person}{Simon Gustavsson}, {and} \bibinfo{person}{William~D Oliver}.} \bibinfo{year}{2020}\natexlab{}.
\newblock \showarticletitle{Superconducting qubits: Current state of play}.
\newblock \bibinfo{journal}{\emph{Annual Review of Condensed Matter Physics}}  \bibinfo{volume}{11} (\bibinfo{year}{2020}), \bibinfo{pages}{369--395}.
\newblock


\bibitem[Krantz et~al\mbox{.}(2019)]%
        {krantz2019quantum}
\bibfield{author}{\bibinfo{person}{Philip Krantz}, \bibinfo{person}{Morten Kjaergaard}, \bibinfo{person}{Fei Yan}, \bibinfo{person}{Terry~P Orlando}, \bibinfo{person}{Simon Gustavsson}, {and} \bibinfo{person}{William~D Oliver}.} \bibinfo{year}{2019}\natexlab{}.
\newblock \showarticletitle{A quantum engineer's guide to superconducting qubits}.
\newblock \bibinfo{journal}{\emph{Applied Physics Reviews}} \bibinfo{volume}{6}, \bibinfo{number}{2} (\bibinfo{year}{2019}), \bibinfo{pages}{021318}.
\newblock
\urldef\tempurl%
\url{https://doi.org/10.1063/1.5089550}
\showDOI{\tempurl}


\bibitem[L'Abbate et~al\mbox{.}(2024)]%
        {l2024quantum}
\bibfield{author}{\bibinfo{person}{Ryan L'Abbate}, \bibinfo{person}{Anthony D'Onofrio}, \bibinfo{person}{Samuel Stein}, \bibinfo{person}{Samuel Yen-Chi Chen}, \bibinfo{person}{Ang Li}, \bibinfo{person}{Pin-Yu Chen}, \bibinfo{person}{Juntao Chen}, {and} \bibinfo{person}{Ying Mao}.} \bibinfo{year}{2024}\natexlab{}.
\newblock \showarticletitle{A quantum-classical collaborative training architecture based on quantum state fidelity}.
\newblock \bibinfo{journal}{\emph{IEEE Transactions on Quantum Engineering}}  \bibinfo{volume}{5} (\bibinfo{year}{2024}), \bibinfo{pages}{1--14}.
\newblock


\bibitem[Li et~al\mbox{.}(2025)]%
        {li2025quantum}
\bibfield{author}{\bibinfo{person}{Guodong Li}, \bibinfo{person}{Fangce Yu}, \bibinfo{person}{Qingle Wang}, \bibinfo{person}{Lin Liu}, \bibinfo{person}{Ying Mao}, {and} \bibinfo{person}{Long Cheng}.} \bibinfo{year}{2025}\natexlab{}.
\newblock \showarticletitle{Quantum neural network classifier with differential privacy}.
\newblock \bibinfo{journal}{\emph{Physica Scripta}} \bibinfo{volume}{100}, \bibinfo{number}{3} (\bibinfo{year}{2025}), \bibinfo{pages}{035109}.
\newblock


\bibitem[Liu et~al\mbox{.}(2024)]%
        {liu2024training}
\bibfield{author}{\bibinfo{person}{Chen-Yu Liu}, \bibinfo{person}{En-Jui Kuo}, \bibinfo{person}{Chu-Hsuan~Abraham Lin}, \bibinfo{person}{Sean Chen}, \bibinfo{person}{Jason~Gemsun Young}, \bibinfo{person}{Yeong-Jar Chang}, {and} \bibinfo{person}{Min-Hsiu Hsieh}.} \bibinfo{year}{2024}\natexlab{}.
\newblock \showarticletitle{Training classical neural networks by quantum machine learning}. In \bibinfo{booktitle}{\emph{2024 IEEE International Conference on Quantum Computing and Engineering (QCE)}}, Vol.~\bibinfo{volume}{2}. IEEE, \bibinfo{pages}{34--38}.
\newblock


\bibitem[Mirrahimi et~al\mbox{.}(2014)]%
        {mirrahimi2014cat}
\bibfield{author}{\bibinfo{person}{Mazyar Mirrahimi}, \bibinfo{person}{Zaki Leghtas}, \bibinfo{person}{Victor~V Albert}, \bibinfo{person}{Steven Touzard}, \bibinfo{person}{Robert~J Schoelkopf}, \bibinfo{person}{Liang Jiang}, {and} \bibinfo{person}{Michel~H Devoret}.} \bibinfo{year}{2014}\natexlab{}.
\newblock \showarticletitle{Dynamically protected cat-qubits: A new paradigm for universal quantum computation}.
\newblock \bibinfo{journal}{\emph{New Journal of Physics}} \bibinfo{volume}{16}, \bibinfo{number}{4} (\bibinfo{year}{2014}), \bibinfo{pages}{045014}.
\newblock
\urldef\tempurl%
\url{https://doi.org/10.1088/1367-2630/16/4/045014}
\showURL{%
\tempurl}


\bibitem[Monroe and et~al.(2014)]%
        {monroe2014large}
\bibfield{author}{\bibinfo{person}{Christopher Monroe} {and} \bibinfo{person}{et al.}} \bibinfo{year}{2014}\natexlab{}.
\newblock \showarticletitle{Large-scale modular quantum-computer architecture with atomic memory and photonic interconnects}.
\newblock \bibinfo{journal}{\emph{Physical Review A}} \bibinfo{volume}{89}, \bibinfo{number}{2} (\bibinfo{year}{2014}), \bibinfo{pages}{022317}.
\newblock


\bibitem[Mu et~al\mbox{.}(2022)]%
        {mu2022iterative}
\bibfield{author}{\bibinfo{person}{Wenrui Mu}, \bibinfo{person}{Ying Mao}, \bibinfo{person}{Long Cheng}, \bibinfo{person}{Qingle Wang}, \bibinfo{person}{Weiwen Jiang}, {and} \bibinfo{person}{Pin-Yu Chen}.} \bibinfo{year}{2022}\natexlab{}.
\newblock \showarticletitle{Iterative qubits management for quantum index searching in a hybrid system}. In \bibinfo{booktitle}{\emph{2022 IEEE International Performance, Computing, and Communications Conference (IPCCC)}}. IEEE, \bibinfo{pages}{283--289}.
\newblock


\bibitem[Namakshenas et~al\mbox{.}(2024)]%
        {namakshenas2024federated}
\bibfield{author}{\bibinfo{person}{Danyal Namakshenas}, \bibinfo{person}{Abbas Yazdinejad}, \bibinfo{person}{Ali Dehghantanha}, {and} \bibinfo{person}{Gautam Srivastava}.} \bibinfo{year}{2024}\natexlab{}.
\newblock \showarticletitle{Federated quantum-based privacy-preserving threat detection model for consumer internet of things}.
\newblock \bibinfo{journal}{\emph{IEEE Transactions on Consumer Electronics}} \bibinfo{volume}{70}, \bibinfo{number}{3} (\bibinfo{year}{2024}), \bibinfo{pages}{5829--5838}.
\newblock


\bibitem[Neven(2024)]%
        {Google2024Willow}
\bibfield{author}{\bibinfo{person}{Hartmut Neven}.} \bibinfo{year}{2024}\natexlab{}.
\newblock \bibinfo{title}{{Meet Willow, our state-of-the-art quantum chip}}.
\newblock \bibinfo{howpublished}{\url{https://blog.google/technology/research/google-willow-quantum-chip/}}.
\newblock
\newblock
\shownote{Accessed: 2025-04-09}.


\bibitem[Nickerson et~al\mbox{.}(2014)]%
        {nickerson2014freely}
\bibfield{author}{\bibinfo{person}{Naomi~H. Nickerson}, \bibinfo{person}{Joseph~F. Fitzsimons}, {and} \bibinfo{person}{Simon~C. Benjamin}.} \bibinfo{year}{2014}\natexlab{}.
\newblock \showarticletitle{Freely scalable quantum technologies using cells of 5-to-50 qubits with very lossy and noisy photonic links}.
\newblock \bibinfo{journal}{\emph{Nature Communications}}  \bibinfo{volume}{4} (\bibinfo{year}{2014}), \bibinfo{pages}{1756}.
\newblock


\bibitem[Nielsen and Chuang(2010)]%
        {nielsen2010quantum}
\bibfield{author}{\bibinfo{person}{Michael~A Nielsen} {and} \bibinfo{person}{Isaac~L Chuang}.} \bibinfo{year}{2010}\natexlab{}.
\newblock \bibinfo{booktitle}{\emph{Quantum Computation and Quantum Information}}.
\newblock \bibinfo{publisher}{Cambridge University Press}.
\newblock


\bibitem[Norris et~al\mbox{.}(2025)]%
        {Norris2025}
\bibfield{author}{\bibinfo{person}{Graham~J. Norris} {et~al\mbox{.}}} \bibinfo{year}{2025}\natexlab{}.
\newblock \showarticletitle{Performance Characterization of a Multi-Module Quantum Processor with Static Inter-Chip Couplers}.
\newblock \bibinfo{journal}{\emph{arXiv preprint arXiv:2503.12603}} (\bibinfo{date}{March} \bibinfo{year}{2025}).
\newblock
\urldef\tempurl%
\url{https://arxiv.org/abs/2503.12603}
\showURL{%
\tempurl}


\bibitem[Qu et~al\mbox{.}(2024)]%
        {qu2024quantum}
\bibfield{author}{\bibinfo{person}{Zhiguo Qu}, \bibinfo{person}{Lailei Zhang}, {and} \bibinfo{person}{Prayag Tiwari}.} \bibinfo{year}{2024}\natexlab{}.
\newblock \showarticletitle{Quantum fuzzy federated learning for privacy protection in intelligent information processing}.
\newblock \bibinfo{journal}{\emph{IEEE Transactions on Fuzzy Systems}} \bibinfo{volume}{33}, \bibinfo{number}{1} (\bibinfo{year}{2024}), \bibinfo{pages}{278--289}.
\newblock


\bibitem[Rached et~al\mbox{.}(2024)]%
        {BenRached2024}
\bibfield{author}{\bibinfo{person}{Sahar~Ben Rached} {et~al\mbox{.}}} \bibinfo{year}{2024}\natexlab{}.
\newblock \showarticletitle{Benchmarking Emerging Cavity-Mediated Quantum Interconnect Technologies for Modular Quantum Computers}.
\newblock \bibinfo{journal}{\emph{arXiv preprint arXiv:2407.15651}} (\bibinfo{date}{July} \bibinfo{year}{2024}).
\newblock
\urldef\tempurl%
\url{https://arxiv.org/abs/2407.15651}
\showURL{%
\tempurl}


\bibitem[Raghavan et~al\mbox{.}(2025)]%
        {IBMResearch2025AnnualLetter}
\bibfield{author}{\bibinfo{person}{Sriram Raghavan}, \bibinfo{person}{Mukesh Khare}, {and} \bibinfo{person}{Jay Gambetta}.} \bibinfo{year}{2025}\natexlab{}.
\newblock \bibinfo{title}{The 2024 IBM Research Annual Letter}.
\newblock \bibinfo{howpublished}{\url{https://research.ibm.com/blog/research-annual-letter-2024}}.
\newblock
\newblock
\shownote{Accessed: 2025-04-09}.


\bibitem[Stein et~al\mbox{.}(2021a)]%
        {stein2021qugan}
\bibfield{author}{\bibinfo{person}{Samuel~A Stein}, \bibinfo{person}{Betis Baheri}, \bibinfo{person}{Daniel Chen}, \bibinfo{person}{Ying Mao}, \bibinfo{person}{Qiang Guan}, \bibinfo{person}{Ang Li}, \bibinfo{person}{Bo Fang}, {and} \bibinfo{person}{Shuai Xu}.} \bibinfo{year}{2021}\natexlab{a}.
\newblock \showarticletitle{Qugan: A quantum state fidelity based generative adversarial network}. In \bibinfo{booktitle}{\emph{2021 IEEE international conference on quantum computing and engineering (QCE)}}. IEEE, \bibinfo{pages}{71--81}.
\newblock


\bibitem[Stein et~al\mbox{.}(2022)]%
        {stein2022quclassi}
\bibfield{author}{\bibinfo{person}{Samuel~A Stein}, \bibinfo{person}{Betis Baheri}, \bibinfo{person}{Daniel Chen}, \bibinfo{person}{Ying Mao}, \bibinfo{person}{Qiang Guan}, \bibinfo{person}{Ang Li}, \bibinfo{person}{Shuai Xu}, {and} \bibinfo{person}{Caiwen Ding}.} \bibinfo{year}{2022}\natexlab{}.
\newblock \showarticletitle{Quclassi: A hybrid deep neural network architecture based on quantum state fidelity}.
\newblock \bibinfo{journal}{\emph{Proceedings of Machine Learning and Systems}}  \bibinfo{volume}{4} (\bibinfo{year}{2022}), \bibinfo{pages}{251--264}.
\newblock


\bibitem[Stein et~al\mbox{.}(2021b)]%
        {stein2021hybrid}
\bibfield{author}{\bibinfo{person}{Samuel~A Stein}, \bibinfo{person}{Ryan L’Abbate}, \bibinfo{person}{Wenrui Mu}, \bibinfo{person}{Yue Liu}, \bibinfo{person}{Betis Baheri}, \bibinfo{person}{Ying Mao}, \bibinfo{person}{Guan Qiang}, \bibinfo{person}{Ang Li}, {and} \bibinfo{person}{Bo Fang}.} \bibinfo{year}{2021}\natexlab{b}.
\newblock \showarticletitle{A hybrid system for learning classical data in quantum states}. In \bibinfo{booktitle}{\emph{2021 IEEE International Performance, Computing, and Communications Conference (IPCCC)}}. IEEE, \bibinfo{pages}{1--7}.
\newblock


\bibitem[Technologies(2025)]%
        {Xanadu2025}
\bibfield{author}{\bibinfo{person}{Xanadu~Quantum Technologies}.} \bibinfo{year}{2025}\natexlab{}.
\newblock \showarticletitle{Scaling and Networking a Modular Photonic Quantum Computer}.
\newblock \bibinfo{journal}{\emph{PubMed}} (\bibinfo{date}{February} \bibinfo{year}{2025}).
\newblock
\urldef\tempurl%
\url{https://pubmed.ncbi.nlm.nih.gov/39843755/}
\showURL{%
\tempurl}


\bibitem[Ullah and Garcia-Zapirain(2024)]%
        {ullah2024quantum}
\bibfield{author}{\bibinfo{person}{Ubaid Ullah} {and} \bibinfo{person}{Begonya Garcia-Zapirain}.} \bibinfo{year}{2024}\natexlab{}.
\newblock \showarticletitle{Quantum machine learning revolution in healthcare: a systematic review of emerging perspectives and applications}.
\newblock \bibinfo{journal}{\emph{IEEE Access}}  \bibinfo{volume}{12} (\bibinfo{year}{2024}), \bibinfo{pages}{11423--11450}.
\newblock


\bibitem[Vazquez et~al\mbox{.}(2024)]%
        {Vazquez2024}
\bibfield{author}{\bibinfo{person}{Almudena~Carrera Vazquez} {et~al\mbox{.}}} \bibinfo{year}{2024}\natexlab{}.
\newblock \showarticletitle{Combining Quantum Processors with Real-Time Classical Communication}.
\newblock \bibinfo{journal}{\emph{Nature}}  \bibinfo{volume}{615} (\bibinfo{date}{November} \bibinfo{year}{2024}), \bibinfo{pages}{715--720}.
\newblock
\urldef\tempurl%
\url{https://doi.org/10.1038/s41586-024-08178-2}
\showDOI{\tempurl}


\bibitem[Verma et~al\mbox{.}(2025)]%
        {verma2025quantum}
\bibfield{author}{\bibinfo{person}{Vandana~Rani Verma}, \bibinfo{person}{Dinesh~Kumar Nishad}, \bibinfo{person}{Vishnu Sharma}, \bibinfo{person}{Vinay~Kumar Singh}, \bibinfo{person}{Anshul Verma}, {and} \bibinfo{person}{Dharti~Raj Shah}.} \bibinfo{year}{2025}\natexlab{}.
\newblock \showarticletitle{Quantum machine learning for Lyapunov-stabilized computation offloading in next-generation MEC networks}.
\newblock \bibinfo{journal}{\emph{Scientific Reports}} \bibinfo{volume}{15}, \bibinfo{number}{1} (\bibinfo{year}{2025}), \bibinfo{pages}{405}.
\newblock


\bibitem[Wille et~al\mbox{.}(2023)]%
        {mqtqmap}
\bibfield{author}{\bibinfo{person}{Robert Wille}, \bibinfo{person}{Lukas Burgholzer}, \bibinfo{person}{Stefan Hillmich}, {and} \bibinfo{person}{Nils Quetschlich}.} \bibinfo{year}{2023}\natexlab{}.
\newblock \bibinfo{title}{MQT QMAP: Optimal Qubit Mapping with Noise Adaptation}.
\newblock
\newblock
\urldef\tempurl%
\url{https://github.com/cda-tum/mqt-qmap}
\showURL{%
\tempurl}


\bibitem[Wu et~al\mbox{.}(2024)]%
        {WuModular}
\bibfield{author}{\bibinfo{person}{Xuntao Wu}, \bibinfo{person}{Haoxiong Yan}, \bibinfo{person}{Gustav Andersson}, \bibinfo{person}{Alexander Anferov}, \bibinfo{person}{Ming-Han Chou}, \bibinfo{person}{Christopher~R. Conner}, \bibinfo{person}{Joel Grebel}, \bibinfo{person}{Yash~J. Joshi}, \bibinfo{person}{Shiheng Li}, \bibinfo{person}{Jacob~M. Miller}, \bibinfo{person}{Rhys~G. Povey}, \bibinfo{person}{Hong Qiao}, {and} \bibinfo{person}{Andrew~N. Cleland}.} \bibinfo{year}{2024}\natexlab{}.
\newblock \showarticletitle{Modular Quantum Processor with an All-to-All Reconfigurable Router}.
\newblock \bibinfo{journal}{\emph{Phys. Rev. X}}  \bibinfo{volume}{14} (\bibinfo{date}{Nov} \bibinfo{year}{2024}), \bibinfo{pages}{041030}.
\newblock
Issue 4.
\urldef\tempurl%
\url{https://doi.org/10.1103/PhysRevX.14.041030}
\showDOI{\tempurl}


\bibitem[Zewe(2024)]%
        {MIT2024}
\bibfield{author}{\bibinfo{person}{Adam Zewe}.} \bibinfo{year}{2024}\natexlab{}.
\newblock \showarticletitle{Modular, Scalable Hardware Architecture for a Quantum Computer}.
\newblock \bibinfo{journal}{\emph{MIT News}} (\bibinfo{date}{May} \bibinfo{year}{2024}).
\newblock
\urldef\tempurl%
\url{https://news.mit.edu/2024/modular-scalable-hardware-architecture-quantum-computer-0529}
\showURL{%
\tempurl}


\bibitem[Zewe(2025)]%
        {MIT2025}
\bibfield{author}{\bibinfo{person}{Adam Zewe}.} \bibinfo{year}{2025}\natexlab{}.
\newblock \showarticletitle{Device Enables Direct Communication Among Multiple Quantum Processors}.
\newblock \bibinfo{journal}{\emph{MIT News}} (\bibinfo{date}{March} \bibinfo{year}{2025}).
\newblock
\urldef\tempurl%
\url{https://news.mit.edu/2025/device-enables-direct-communication-among-multiple-quantum-processors-0321}
\showURL{%
\tempurl}


\end{thebibliography}

\end{document}